\documentstyle[preprint,aps,epsfig,amsmath,amssymb]{revtex}

\tightenlines
\draft

\begin{document}

\title{Dynamic structure factor and momentum distribution
of a trapped Bose gas}
   
\author{F.~Zambelli$^a$, L.~Pitaevskii$^{a,b}$, D.M.~Stamper-Kurn$^c$, and
  S.~Stringari$^a$}
\address{$^a$Dipartimento  di Fisica, Universit\`{a} di Trento,}
\address{and Istituto Nazionale per la Fisica della Materia, I-38050 Povo,
Italy}
\address{$^b$Kapitza Institue for Physical Problems, 117334 Moscow, Russia}
\address{$^c$Norman Bridge Laboratory of Physics,}
\address{California Institute of Technology 12-33, Pasadena, CA 91125}
\date{December 6, 1999}

\maketitle

\begin{abstract}
The dynamic structure factor of a trapped Bose-Einstein condensed gas
is investigated at zero temperature in the framework of Bogoliubov
theory. Different values of  momentum transfer are considered,
ranging from the phonon to the single-particle regime. 
Various approximated schemes are discussed, including the
local density approximation, where the system is locally described
as a uniform gas,  and the impulse approximation, where the
response is fixed by the momentum distribution of the condensate. 
A comprehensive approach, based on the eikonal expansion, is
presented. The predictions of theory are successfully compared with
the results of recent two-photon Bragg scattering experiments, both
at low and high momentum transfer. Relevant features of the dynamic
structure factor are also discussed using the formalism of sum rules
and  the concept of scaling. Particular emphasis is given to the
regime of high momentum transfer, where the dynamic structure
factor is sensitive to the behaviour of the order parameter in
momentum space, and some instructive examples showing the
consequence of long-range coherence are presented. 
\end{abstract}

\pacs{PACS numbers: 03.65.-w, 05.30.Jp, 32.80.-t, 67.40.-w}

\narrowtext
  
\begin{section}{Introduction}
The dynamic structure factor provides an important characterization of
the dynamic behaviour of quantum many-body systems.
In particular, its exploration has played a crucial role in
understanding  the physics of  superfluid $^4$He, starting
from the measurement of the roton spectrum 
\cite{Palewski} until the more recent determinations of the condensate 
fraction available from neutron scattering experiments \cite{Sokol}.
The dynamic structure factor is measurable through inelastic
scattering, in which the probe particle is weakly coupled to the many body
system so that the scattering may be described within the
Born approximation \cite{VanHove}. 
In the case of dilute gases it can be measured via inelastic light 
scattering as recently shown by the experiments of Refs. 
\cite{Bragg1,Bragg2} carried out on a trapped Bose gas of sodium atoms.
The dynamic structure factor provides information on both the spectrum
of collective excitations which can be investigated at low momentum
transfer, and the momentum distribution  which characterizes the behaviour
of the system at  high momentum transfer, where the response
is dominated by single-particle effects.

In superfluid helium the typical momentum giving the
transition between the collective and the single-particle behaviour is
fixed by the inverse of the range of two-body forces, a value close to 
the average interatomic distance. 
At larger momenta one explores microscopic features of the system
which are sensitive to short range correlations and to the details of
the two-body interaction.
The situation is very different in a dilute gas where the
transition takes place at momenta much smaller than the inverse of
the scattering length, which fixes the range of interactions.
As a consequence, in a Bose gas one can explore a domain of
relatively high momenta, where the response of the system is not 
affected by collective features, nor by short range correlations, 
but is determined by the momentum distribution of the condensate.
While in a uniform system this distribution is a simple
$\delta$-function, in a trapped gas it exhibits a non trivial
behaviour and is strongly affected by the presence of two-body
interactions.

The purpose of this paper is to provide a systematic theoretical
discussion of the behaviour of the dynamic structure factor of
inhomogeneous Bose-Einstein condensates at zero temperature
and to make quantitative comparisons with the recent data obtained 
with two photon Bragg scattering experiments \cite{Bragg1,Bragg2}, 
pointing out the role of two-body interactions at both low and high
momentum transfer.
The applicability of both the local density (LDA) and of
the impulse approximation (IA) will be discussed in detail. 
The LDA assumes that the system can be locally described as a
uniform gas, and is adequate at moderately low values of momentum transfer.
Conversely, the IA assumes that the dynamic structure factor is
sensitive to the momentum distribution, which, for Bose-Einstein
condensed systems, is determined not locally but globally,
according to the size and shape of the condensate wavefunction.
A comprehensive description of both the LDA and IA regimes will be
presented using the eikonal expansion, which holds
in the single particle regime at high momentum transfer.
Special emphasis will be given to sum rules as well as to the scaling
behaviour exhibited by the dynamic structure factor in the IA regime.
Finally we will point out the occurrence of interesting
features exhibited by the dynamic structure factor in
the presence of vortices and of interference effects in
momentum space.
\end{section}

\begin{section}{Dynamic structure factor and Bogoliubov theory}
The dynamic structure factor of a many body system is defined by the
expression
\begin{equation}
S({\mathbf q},E)=\frac{1}{{\cal Z}}\sum_{mn} e^{-\beta E_m}
|\langle m|\rho_{{\mathbf q}}|n\rangle|^2
\delta (E-E_m+E_n)
\label{sqo}
\end{equation}
where ${\mathbf q}$ and $E$ are the momentum and
energy transferred by the probe to the sample. In Eq. (\ref{sqo})
$|n\rangle$ and $E_n$ are the eigenstates and eigenvalues of
the Hamiltonian of the system, $e^{-\beta E_n}$ is the usual Boltzmann
factor, $\rho_{{\mathbf q}}=\sum_je^{i{\mathbf q}\cdot{\mathbf r}_j/\hslash}$
is the Fourier transform of the one-body density operator, and ${\cal
  Z}$ is the usual canonical partition function.

In Refs. \cite{Bragg1,Bragg2}, the dynamic structure factor of a
trapped Bose-Einstein condensate is measured using two-photon
optical Bragg spectroscopy. Two laser beams are impingent upon the
condensate. The difference in the wavevectors of the beams defines
the momentum transfer ${\mathbf q}$, and the frequency difference
between the beams defines the energy transfer $E$. The atoms exposed
to these beams can undergo a stimulated light scattering event by
absorbing a photon from one of the beams and emitting into the
other. After exposure to these laser beams, the response of the
condensate is measured by a time of flight technique by which the
number of optically excited atoms can be counted. The momentum
transfer in each of these experiments was fixed by the particular
optical setup, while the energy transfer was scanned by varying the
frequency difference between the beams. Because atoms could be
scattered by absorbing a photon from either of the laser beams, the
response of the system actually measures the combination
$S({\mathbf q},E)-S(-{\mathbf q},-E)$, thus significantly
suppressing the effects of temperature in the measured signal.

Let us start our discussion by recalling that in an ideal uniform
gas  Eq.(\ref{sqo}) takes the simple integral form \cite{IG}
\begin{equation}
S({\mathbf q},E)=\int\!d{\mathbf p}\;
n({\mathbf p})[1\pm n({\mathbf p}+{\mathbf q})]
\delta\bigg(E-\frac{({\mathbf p}+{\mathbf q})^2}{2m}+\frac{p^2}{2m}\bigg)\;.
\label{sqoidT}
\end{equation}
where $n({\mathbf p})=\langle a_{{\mathbf p}}^{\dagger}
a_{{\mathbf p}}\rangle$ is the statistical average of the momentum
operator $a_{{\mathbf p}}^{\dagger}a_{{\mathbf p}}$, and the sign
$+\;(-)$ holds for Bose (Fermi) statistics.
The scattering process is hence enhanced and suppressed
in Bose and Fermi gases respectively. In the fermionic case this
reflects the Pauli exclusion principle.
For large momentum transfer ${\mathbf q}$ the exchange term of Eq.
(\ref{sqoidT}) is negligible because the momentum distribution decreases
rapidly at high momenta and one finds the expression
\begin{equation}
S_{IA}({\mathbf q},E)=\int\!d{\mathbf p}\;
\delta\bigg(E-\frac{({\mathbf p}+{\mathbf q})^2}{2m} +\frac{p^2}{2m}\bigg)
n({\mathbf p})\;,
\label{sqoia}
\end{equation}
known as the impulse approximation (IA) \cite{Hohenberg}.
It is important to stress that the validity of Eq.(\ref{sqoia}) is not 
restricted to the ideal gas, but holds in general at high momentum
transfer also for interacting and non uniform systems, independent of 
quantum statistics.
Of course in this case the momentum distribution $n({\mathbf p})$
will differ significantly from that of the ideal gas.

At small momentum transfer the ideal gas prediction (\ref{sqoidT}) is
inadequate, especially in the case of Bose gases, in which
mean field interactions drastically modify the structure of
$S({\mathbf q},E)$, giving rise to the propagation of phonons.

The calculation of $S({\mathbf q},E)$ in interacting many-body systems
requires in general a major theoretical effort.
In the following we will limit ourselves to the $T=0$ case and to
the study of  dilute Bose gases where Bogoliubov theory is applicable.
This restricts the range of momenta $q$ to the ``macroscopic'' regime
$qa\ll\hslash$ where $a$ is the $s$-wave scattering length. 
For larger values of $q$ short range correlations become important and 
Bogoliubov theory is no longer adequate.
In the conditions of the experiment of \cite{Bragg1}, carried out on a
gas of sodium atoms, the Bogoliubov approach is well applicable since
$q/\hslash \sim 20$ $\mu$m$^{-1}$ and  $qa\sim 0.06\;\hslash$.
The macroscopic condition is even better satisfied in the experiment of
\cite{Bragg2} where smaller values of $q$ have been used.

According to Bogoliubov theory the excited states of the system are
given by the solution of the coupled equations \cite{LP}
\begin{align}
\epsilon u({\mathbf r}) &= \bigg[-\frac{\hslash^2}{2m}\nabla^2 +
V_{\rm ext}({\mathbf r})-\mu +2g\psi_0^2({\mathbf r})\bigg] u({\mathbf r})
+ g\psi_0^2({\mathbf r})v({\mathbf r})
\label{Bogu}\\
-\epsilon v({\mathbf r}) &= \bigg[-\frac{\hslash^2}{2m}\nabla^2 +
V_{\rm ext}({\mathbf r})-\mu +2g\psi_0^2({\mathbf r})\bigg] v({\mathbf r})
+ g\psi_0^2({\mathbf r})u({\mathbf r})
\label{Bogv}
\end{align}
for the ``particle'' and ``hole'' components of the elementary modes.
In Eqs. (\ref{Bogu}-\ref{Bogv}) $\epsilon$ is the energy of the
excitation, and $V_{\rm ext}({\mathbf r})$ is the external potential 
for which, unless differently specified, we make the axially symmetric 
harmonic choice:
\begin{equation}
V_{\rm ext}({\mathbf r})=
\frac{1}{2}m\omega_{\perp}^2(x^2+y^2)+\frac{1}{2}m\omega_z^2z^2 \,.
\label{Vext}
\end{equation}
Furthermore $\mu$ is the chemical potential, $g=4\pi\hslash^2a/m$ is the
coupling constant which will be assumed to be positive, and finally
$\psi_0({\mathbf r})$ is the order parameter characterizing the ground
state of the system.

In terms of the functions $u_n({\mathbf r})$ and $v_n({\mathbf r})$, which
satisfy the ortho-normalization condition \cite{Fetter1}
\begin{equation}
\int\!d{\mathbf r}[u_n^{\ast}({\mathbf r})
u_m({\mathbf r})-v_n^{\ast}({\mathbf r})v_m({\mathbf r})]=\delta_{nm}\;,
\label{norm}
\end{equation}
the relevant matrix element of the density operator takes the form
\begin{equation}
\langle n|\rho_{{\mathbf q}}|0\rangle =
\int\!d{\mathbf r}[u_n^{\ast}({\mathbf r})+v_n^{\ast}({\mathbf r})]
e^{i{\mathbf q}\cdot{\mathbf r}/\hslash}\psi_0({\mathbf r})
\label{matrixelement}
\end{equation}
and the dynamic structure factor then becomes, at $T=0$, \cite{Griffin}
\begin{equation}
S({\mathbf q},E)=\sum_n\bigg|\int\!d{\mathbf r}
[u_n^{\ast}({\mathbf r})+v_n^{\ast}({\mathbf r})]e^{i{\mathbf q}
\cdot{\mathbf r}/\hslash}\psi_0({\mathbf r})\bigg|^2\delta (E-\epsilon_n)\;.
\label{sqobog}
\end{equation}

In the case of a uniform gas the ``particle'' and ``hole'' components are
plane waves: $u({\mathbf r})=U\exp{(i{\mathbf p}\cdot{\mathbf r}/\hslash)}$
and $v({\mathbf r})=V\exp{(i{\mathbf p}\cdot{\mathbf r}/\hslash)}$ and the
coupled Equations (\ref{Bogu}-\ref{Bogv}) give rise to the Bogoliubov
excitation spectrum \cite{Bogoliubov}:
\begin{equation}
\epsilon(p)=\sqrt{\frac{p^2}{2m}\bigg(\frac{p^2}{2m}+2gn\bigg)}\;,
\label{Bogspectrum}
\end{equation}
where $n$ is the density of the gas.
This spectrum exhibits a phonon dispersion $\epsilon =cp$ at low momenta,
with the velocity of sound given by $c=\sqrt{gn/m}$, while in the
opposite limit of high momenta it approaches the free particle energy
$\epsilon = p^2/2m$.
The transition between the collective and the single particle
behaviour occurs at momenta of the order of $\hslash/\xi$, where
\begin{equation}
\xi =\frac{1}{\sqrt{8\pi na}}
\label{healing}
\end{equation}
is the so-called healing length.

Using the normalization condition (\ref{norm}), one  finds the result
\begin{equation}
S_B({\mathbf q},E)=N\frac{q^2}{2m\epsilon(q)}\,\delta(E-\epsilon(q))
\label{sqobogom}
\end{equation}
for the dynamic structure factor, which consists of a $\delta$-function
centered at the Bogoliubov frequency (\ref{Bogspectrum}).
Eq. (\ref{sqobogom}) yields the Feynman-like result
\begin{equation}
S_B({\mathbf q})=\frac{1}{N}\int\!dE S({\mathbf q},E)=
\frac{q^2}{2m\epsilon(q)}
\label{m0om}
\end{equation}
for the static structure factor which tends linearly to zero at low
momenta, and saturates to 1 in the opposite, high $q$ limit.

Results (\ref{sqobogom}-\ref{m0om}) hold for uniform Bose gases. 
In the presence of non uniform trapping a natural generalization is 
provided by the local density approximation (LDA), according to which 
the system behaves locally as a piece of uniform gas whose dynamic 
structure factor is given by the expression (\ref{sqobogom}),
evaluated at the corresponding density \cite{Timmermans}:
\begin{equation}
S_{LDA}({\mathbf q},E)=\int\!d{\mathbf r}\;n({\mathbf r})\delta
(E-\epsilon ({\mathbf r},{\mathbf q}))\frac{q^2}{2m
\epsilon ({\mathbf r},{\mathbf q})}\;.
\label{sqold}
\end{equation}
In Eq. (\ref{sqold}) $n({\mathbf r})$ is the ground state density
of the system, and $\epsilon ({\mathbf r},{\mathbf q})=\epsilon
(n({\mathbf r}),{\mathbf q})$ is the local Bogoliubov dispersion
(\ref{Bogspectrum}).
Eq. (\ref{sqold}) is expected to describe accurately the dynamic structure
factor for momenta larger than $\hslash/R$ where $R$ is the radius of the
condensate, since the effects of discretization in the
excitation spectrum can be safely ignored.
At the same time the momentum transfer $q$ should not be too large
since the local density approximation ignores the Doppler effect
associated with the spreading of the momentum distribution of the
condensate, which is expected to become the leading effect in the
dynamic structure factor at very large values of $q$.
The Doppler broadening is accounted for by the impulse
approximation (\ref{sqoia}), which, however, ignores the mean field effects
of Bogoliubov theory.
The conditions of applicability for both LDA and IA will be
established in the next section.
\end{section}

\begin{figure}
\begin{center}
\input{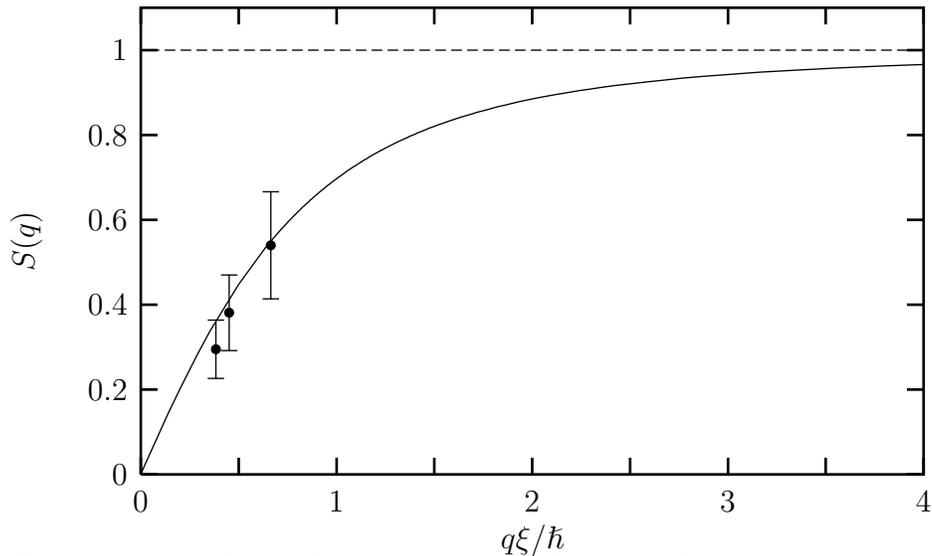}
\caption{Static structure factor $S(q)$ at $T=0$ as a 
function of $q\xi/\hslash$. 
The prediction (\ref{Sqdan}) of the local density approximation (solid
line) is compared with the experimental points taken from 
\protect\cite{Bragg2}. The result of the non-interacting model 
(dashed line) is also indicated.}
\label{fig1}
\end{center}
\end{figure}

\begin{figure}
\begin{center}
\input{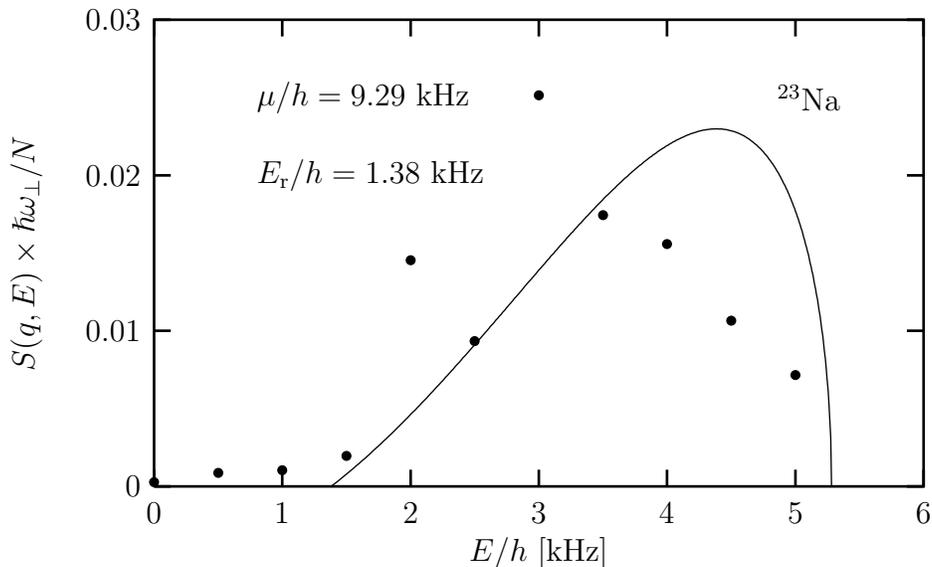}
\caption{Dynamic structure factor calculated
using the local density approximation (\ref{sqoldTF}). 
Experimental points are taken from Ref. \protect\cite{Bragg2}.
The trapping frequencies are $\omega_{\perp}=2\pi\,150$ Hz and 
$\omega_z=2\pi\,18$ Hz.}
\label{fig2}
\end{center}
\end{figure}

\begin{section}{Local density, impulse and eikonal approximations}
In the previous section we introduced two useful
approximations to the dynamic structure factor of an interacting
Bose gas: the impulse (\ref{sqoia}) and local density
(\ref{sqold}) approximations. These two descriptions hold in
different regimes of momentum transfer. The purpose of this
section is to discuss the corresponding predictions
and conditions of applicability. We will also present a
comprehensive description of the high $q$ response of the system,
based on eikonal expansion, which includes the LDA and the IA as
special cases.

\begin{subsection}{Local Density Approximation}
Let us first discuss the local density approximation (\ref{sqold}).
An explicit expression for $S({\mathbf q},E)$ can be obtained working
in the Thomas-Fermi limit $Na/a_{\rm ho}\gg 1$, where the
ground state density is given by (see, for example, \cite{rmp})
\begin{equation}
n({\mathbf r})=\frac{1}{g}\big(\mu -V_{\rm ext}({\mathbf r})\big)\;,
\label{TFdensity}
\end{equation}
and the chemical potential takes the form
\begin{equation}
\mu=\bigg(15\frac{Na}{a_{{\rm ho}}}\bigg)^{2/5}\hslash\omega_{{\rm ho}}\;.
\label{muTF}
\end{equation}
In Eq. (\ref{muTF}) $a_{{\rm ho}}=\sqrt{\hslash/m\omega_{\rm ho}}$ 
is the oscillator length calculated using the geometrical average 
$\omega_{{\rm ho}}=
(\omega_{\perp}^2\omega_z)^{1/3}$ of the oscillator frequencies.
Using this density profile and the
result of Eq. (\ref{Bogspectrum}) for the excitation spectrum, one obtains
$S({\mathbf q}, E)$ for a trapped Bose condensate as \cite{Timmermans,Bragg2}:
\begin{equation}
S_{\rm LDA}({\mathbf q},E)=\frac{15}{8}\frac{(E^2-E^2_{\rm r})}{E_{\rm r}\mu^2}
\sqrt{1-\frac{(E^2-E_{\rm r}^2)}{2E_{\rm r}\mu}}\;,
\label{sqoldTF}
\end{equation}
where
\begin{equation}
E_{\rm r}=\frac{q^2}{2m}
\label{recoil}
\end{equation}
is the free recoil energy.
Differently from the case of a uniform gas (see Eq.(\ref{sqobogom})),
the dynamic structure factor is no longer a $\delta$-function, its
value being different from zero in the interval
$E_{\rm r} < E < E_{\rm r}\sqrt{1+2\mu/E_{\rm r}}$.
The value $E=E_{\rm r}$ corresponds to the excitation energy in the
region near the border where the gas is extremely dilute and hence
non-interacting. The value $E=E_{\rm r}\sqrt{1+2\mu/E_{\rm r}}$
is the excitation energy of a Bogoliubov gas evaluated at the
central density.
Notice that the LDA expression (\ref{sqoldTF}) for $S({\mathbf q},E)$
does not depend on the direction of the vector ${\mathbf q}$ even in
the presence of a deformed trap. 

As pointed out in the previous section, the local density
approximation requires that the momentum transfer be larger than
$\hslash/R$. For smaller values of $q$ the response of
the system is sensitive to the discretized modes of the system and
the LDA cannot be longer employed.
The theoretical analysis of the dynamic structure factor in this
regime of low momentum transfer has been carried out in \cite{Griffin}.
In the following we will always assume that the condition
$q\gg\hslash/R$ is satisfied.

Starting from (\ref{sqoldTF}) one can evaluate the inelastic
static structure factor
$S({\bf q})= N^{-1} \int
S({\mathbf q},E)\;dE$, which takes the analytic form
\cite{Bragg2}:
\begin{equation}
S(q)=\frac{15}{4}\left\{\frac{3+\alpha}{4\alpha^2}-
\frac{(3+2\alpha-\alpha^2)}{16\alpha^{5/2}}\left[\pi+2\arctan
{\left(\frac{\alpha
-1}{2\sqrt{\alpha}}\right)}\right]\right\}\;,
\label{Sqdan}
\end{equation}
with $\alpha=2\mu/E_{\rm r}$. When $q\to 0$ Eq. (\ref{Sqdan}) vanishes 
as $S(q)\sim q/(2m\bar{c})$, reflecting the
role played by dynamic correlations which strongly suppress light
scattering in the phonon regime (see Fig. \ref{fig1}). 
Here $\bar{c}=32\sqrt{\mu/m}/15\pi$
corresponds to an average sound velocity, which, as expected, is
smaller than the value $\sqrt{\mu /m}$ calculated in the center of the
trap.

Useful information about the dynamic structure factor can be obtained
by evaluating its energy moments
$m_k({\mathbf q})=\int_{0^+}^{\infty}\!E^kS({\mathbf q},E)dE$.
For example the average excitation energy $\bar{E}$ and the rms
width $\Delta_{\rm rms}$ can be defined as
\begin{equation}
\bar{E}=\frac{m_1}{m_0}\;,
\label{m1/m0}
\end{equation}
and 
\begin{equation}
\Delta_{\rm rms}=\sqrt{\frac{m_2}{m_0}-\bigg(\frac{m_1}{m_0}\bigg)^2}
\label{width}
\end{equation}
respectively.

Let us first discuss the phonon regime in which $E_{\rm r}\ll\mu$, or,
equivalently, $q\xi\ll\hslash$, where $\xi$ is the healing length
(\ref{healing}) calculated at the central density. 
Typical values of $\xi$ in sodium samples, where $a = 2.75$ nm,  are of
$0.1\div 1$ $\mu$m depending on the density of the gas. Using the
LDA expression (\ref{sqoldTF}) for $S({\mathbf q},E)$ one finds, in
the phonon regime, the results  $\bar{E}=\bar{c}q$ and
$\Delta_{\rm rms}\simeq 0.3\;\bar{c}q$, showing that the width of
the signal is not much smaller than the average energy.
The shape of $S({\bf q}, E)$ turns out to be asymmetric 
as a function of $E$ (see Fig. \ref{fig2}), and the peak response 
occurs at an energy which is higher (by about 15\%) than $\bar{E}$.

The dynamic structure factor $S({\mathbf q},E)$ has been recently
measured in the phonon regime \cite{Bragg2}. 
A typical experimental curve is reported in Fig. \ref{fig2}
together with the prediction (\ref{sqoldTF}).
These measurements also show (see Fig. \ref{fig1}) the static structure factor
$S(q)$ in the phonon regime to be smaller than the non interacting gas
value $S(q)=1$, in agreement with the predictions of Eq. (\ref{Sqdan})
\cite{notaIG}.

In the opposite limit of large momentum transfer, where $q\xi\gg\hslash$,
the excitation energy (\ref{m1/m0}) predicted by the LDA is given by the
expression
\begin{equation}
\bar{E}= E_{\rm r} + \frac{4}{7}\mu\;.
\label{shift}
\end{equation}
This result corresponds to the average of the Bogoliubov energy
$\epsilon({\mathbf q},{\mathbf r})=q^2/2m+gn({\mathbf r})$ holding at
high $q$. Notice that, due to the asymmetric shape of the dynamic structure
factor (\ref{sqoldTF}), the average energy (\ref{shift}) turns out to
be smaller than the peak energy
\begin{equation}
E_{\rm peak}=E_{\rm r}+\frac{2}{3}\mu\;.
\label{epeak}
\end{equation}
This asymmetry should be, in principle, taken into account in the fit
of experimental data. 
However, the shift of the line center was effectively determined from 
experiments in \cite{Bragg1} using symmetric Gaussian fits and 
was in good agreement with Eq. (\ref{shift}) (see Fig. \ref{fig3}). In the
same regime of large $q$ the rms width predicted by the LDA is given by
\begin{equation}
\Delta_{\rm LDA}=\sqrt{\frac{8}{147}}\;\mu
\label{Deltahighq}
\end{equation}
and is independent of $q$.

Thus, for momentum transfers $q$ which are sufficiently small to
allow the use of the local density approximation, the structure
factor of an inhomogeneous Bose-Einstein condensate can be derived
from the Bogoliubov spectrum for a uniform condensed gas.  Studies
in this regime therefore serve as a probe of the spectrum of both
collective ($q \xi \ll \hslash$) and free-particle ($q \xi \gg
\hslash$) excitations. 
\end{subsection}

\begin{subsection}{Impulse Approximation}
Let us now discuss the response
of a trapped condensate to very large momentum transfers at which
the form of the dynamic structure factor is dominated by Doppler
broadening. In this regime, the dynamic structure factor is
correctly described by the impulse approximation (\ref{sqoia}).
Inelastic scattering at such high momentum transfers allows one to
directly measure the momentum distribution of a trapped Bose gas.
The possibility of such measurements is highly appealing since
most of experimental investigations in these systems have been so
far limited to the study of density profiles. In current
experiments on  harmonically-confined Bose gases, the sizes of the
condensate $R$ and of the thermal cloud $R_T$ are typically
comparable. In the Thomas-Fermi regime, the ratio between the two
radii is given as
\begin{equation}
\frac{R}{R_T} \sim \sqrt{\frac{\mu}{k_B T}} =
\sqrt{\frac{\hslash\omega_{\rm ho}}{k_BT}}\bigg(15\frac{N_0a}{a_{\rm
ho}} \bigg)^{1/5}\;, \label{rratio}
\end{equation}
where  $N_0$ is the number of atoms in the condensate. Due to
the large value of the Thomas-Fermi parameter $N_0 a/a_{\rm ho}$,
this ratio is typically close to unity. Expression (\ref{rratio})
also provides an estimate for the ratio of sizes of the two
components measured in time-of-flight experiments, in which the
trap is suddenly switched off and the gas allowed to freely
expand.  While the expansion of the thermal cloud is indicative of
the non-condensate momentum distribution before release from the
trap, the expansion of the condensate in the Thomas-Fermi regime
is dominated by the release of interaction energy and does not
reveal its initial momentum distribution.

In contrast, the distinction between the condensate and the
thermal cloud in momentum space is stark.  
A confined condensate of finite size has a momentum distribution of
width $\Delta p_c\sim\hslash/R$ fixed by the inverse of the size $R$ 
of the condensate.
The momentum width of the thermal cloud is instead given by
the temperature of the gas as $\Delta p_T\sim\sqrt{mk_BT}$. For
harmonic confinement in the Thomas-Fermi regime, one then finds
\begin{equation}
\frac{\Delta p_c}{\Delta p_T}\sim \sqrt{\frac{\hslash\omega_{\rm
ho}}{k_BT}}\bigg(15\frac{N_0a}{a_{\rm ho}} \bigg)^{-1/5}\;.
\label{pratio}
\end{equation}
In contrast with the comparison of the condensate and the thermal
cloud in coordinate space (\ref{rratio}), the distinction between
the two components in momentum space is strongly {\em enhanced} by
two body interactions as the Thomas-Fermi parameter $N_0a/a_{\rm
ho}$ increases.  The investigation of the momentum distribution
consequently provides a deeper understanding of the phenomenon of
BEC. In particular the smallness of the width $\Delta p_c$ reflects 
the presence of long-range coherence.
Measuring the momentum distribution at
zero temperature would, in principle, give access also to the quantum
depletion of the condensate. In practice, however, the quantum
depletion is too small and broadened over too large a momentum
range \cite{Huang}  to be observable in present experiments.

The expression (\ref{sqoia}) for the impulse approximation can be also 
written in the form
\begin{equation}
S_{\rm IA}({\mathbf q},E)=\frac{m}{q}\int\!dp_ydp_z\;
n\big(p_x,p_y,p_z\big)\;,
\label{sqoia2}
\end{equation}
where we have assumed that the vector ${\mathbf q}$ is oriented
along the $x$ axis, and $p_x=m(E-E_{\rm r})/q$. The integral
$\int\!dp_ydp_z\;n(p_x,p_y,p_z)$ is also called the longitudinal
momentum distribution. Eqs. (\ref{sqoia}) and (\ref{sqoia2}) show
that in the regime of applicability of the IA one can
extract useful information on the momentum distribution starting
from the experimental measurement of the dynamic structure factor.

In a dilute Bose gas at zero temperature the momentum distribution is 
given by $n({\mathbf p})=|\phi({\mathbf p})|^2$, where
\begin{equation}
\phi({\mathbf p})=(2\pi\hslash)^{-3/2}\int\!d{\mathbf r}\,\psi({\mathbf r})\,
e^{i{\mathbf p}\cdot{\mathbf r}/\hslash}
\label{Fourier}
\end{equation}
is the Fourier transform of the order parameter.
The form of $n({\mathbf p})$ for a trapped condensate has been
discussed previously \cite{Baym,Franco}. In the Thomas-Fermi limit
$Na/a_{{\rm ho}}\gg 1$ one finds the simple analytic result
\begin{equation}
n_{\rm TF}({\mathbf p})=N\frac{15}{16\lambda}\left(\frac{R_{\perp}}
{\hslash}\right)^3
\left[\frac{J_2(\tilde{p})}{\tilde{p}^2}\right]^2\;,
\label{npTF}
\end{equation}
where $J_2(z)$ is the usual Bessel function of order 2,
\begin{equation}
R_{\perp}=\left(15\frac{Na}{a_{\rm ho}}\right)^{1/5}\lambda^{1/3}a_{\rm ho}
\label{RTF}
\end{equation}
is the Thomas-Fermi radius of the condensate in the $x-y$ plane, and
$\tilde{p}=\sqrt{p_x^2+p_y^2+(p_z/\lambda)^2}
R_{\perp}/\hslash$ is a dimensionless variable,
with the parameter $\lambda=\omega_z/\omega_{\perp}$
fixing the anisotropy of the external potential.

Eq. (\ref{npTF}) explicitly shows that the momentum distribution
scales as $1/R_{\perp}$ and is consequently much narrower than 
that of the non-interacting gas
\begin{equation}
n_{\rm IBG}({\mathbf p})=N\left(\frac{a_{\rm ho}}{\hslash\sqrt{\pi}}\right)^3
\exp{\left[-\frac{a_{\rm ho}^2}{\hslash^2}\lambda^{1/3}
\left(p_x^2+p_y^2+\frac{p_z^2}{\lambda}\right)\right]}\;,
\label{npHO}
\end{equation}
since $R_{\perp}\gg a_{\rm ho}$.

In the impulse approximation (\ref{sqoia2}) the peak of
$S({\mathbf q},E)$  coincides with the recoil energy $E_{{\rm r}}$
while the curve is broadened due to the Doppler effect in the momentum
distribution.
A useful estimate of the broadening can be obtained carrying out a
Gaussian expansion in the dynamic structure factor (\ref{sqoia2}) near
the peak value $E=E_{\rm r}$. One finds:
\begin{equation}
S_{\rm IA}({\mathbf q},E)\simeq 
S_{\rm IA}({\mathbf q},E_{\rm r})
\exp{\left[-\frac{(E-E_{\rm r})^2}{2\Delta_{\rm IA}^2}\right]}\;,
\label{exp}
\end{equation}
with $\Delta_{\rm IA}^2 = - [S_{\rm IA}({\mathbf q},E)/
\partial^2_{\rm E}S_{\rm IA}({\mathbf q},E)]|_{\rm E=E_{\rm r}}$.
By calculating the second derivative of (\ref{sqoia2})
with the Thomas-Fermi profile (\ref{npTF}) for the momentum
distribution, we obtain, after some straightforward algebra,
the result
\begin{equation}
\Delta_{\rm IA}=\sqrt{\frac{8}{3}}
\frac{q\hslash}{mR_{\perp}}\;.
\label{Gwidth2}
\end{equation}
The Gaussian profile (\ref{exp}) reproduces very well the exact
curve (see Fig. \ref{fig6}), so that the Doppler width
(\ref{Gwidth2}) can be usefully compared with experiments, where
the widths are usually extracted through Gaussian fits to the
measured signal. The Doppler width (\ref{Gwidth2}) is linear in
$q$, and for large momentum transfer it can become comparable
or even larger than the mean field width (\ref{Deltahighq}). In
Fig. \ref{fig4} we show the theoretical prediction (\ref{Gwidth2})
together with the experimental values obtained at several
densities. This figure confirms that the IA accounts
for the observed widths in the low density regime. At higher
density the mean field effect (\ref{Deltahighq}) can no
longer be neglected.

It is worth noticing that the width (\ref{Gwidth2}) should not be
confused with the rms width (\ref{width}) which
requires the evaluation of the $m_2$ moment and,
using the IA expression (\ref{sqoia}) or (\ref{sqoia2}), takes the
form:
\begin{equation}
\Delta_{\rm rms} =q\sqrt{\frac{2}{m} E_{\rm kin}^x}\;,
\label{Deltadoppler}
\end{equation}
instead of (\ref{Gwidth2}). Here $E_{{\rm
kin}}^x=\int\!d{\mathbf p}\;p_x^2n({\mathbf p})/2m$ is the
$x$-component of the kinetic energy of the condensate. The
evaluation of the kinetic energy requires a careful analysis
\cite{Franco2,Lundh} of the region near the boundary of the
condensate, and cannot be evaluated using the
Thomas-Fermi expression (\ref{npTF}) for $n({\mathbf p})$, which 
incorrectly yields a divergent
result. For large $N$ samples one finds \cite{Franco2}, assuming
isotropic trapping, 
\begin{equation}
\frac{E_{\rm kin}^x}{N}\simeq\frac{5\hslash^2}{6mR^2}
\log{\left(\frac{R}{1.3a_{\rm ho}}\right)}\;.
\label{EkinTF}
\end{equation}
The logarithmic term reflects the fact that the $m_2$ moment, and
hence the rms width, is sensitive to the high energy tails of the
dynamic structure factor, a region which is difficult to
measure since the intensity of the signal in
the tails is very small. Because of this, estimate
(\ref{Gwidth2}) is much more significant from the experimental
point of view than expression (\ref{Deltadoppler}).

The investigation of the dynamic structure factor 
also provides information on the coherence effects
exhibited by the system and in particular on the behaviour of
the off-diagonal one-body density
\begin{multline}
\rho^{(1)}(\textbf{s})=
N\int\!d{\mathbf R}d{\mathbf r}_2 ...d{\mathbf r}_N
\psi^*\left({\mathbf R}+\frac{\textbf{s}}{2},{\mathbf
    r}_2,...,{\mathbf r}_N\right)
\psi\left({\mathbf R}-\frac{\textbf{s}}{2},{\mathbf r}_2,...,{\mathbf
  r}_N\right)\\
=\int\!d{\mathbf p}\,n({\mathbf p})\exp{\left[-i\frac{
{\mathbf p}\cdot{\mathbf s}}{\hslash}\right]}\;,
\label{rho1}
\end{multline}
where $\psi({\mathbf r}_1,...,{\mathbf r}_N)$ is
the many-body wave function of the system, and $n({\mathbf p})$ is the
momentum distribution. 
By taking the Fourier transform of Eq.
(\ref{sqoia2}) with respect to $p_x$, one finds the result
\begin{equation}
\rho^{(1)}(s_x,0,0)=\int\!dE\,S_{\rm IA}({\mathbf q},E)\exp{\left[
-i\frac{ms_x}{\hslash q}(E-E_{\rm r})\right]}\;,
\label{cohe1}
\end{equation}
which shows that the one-body density is a measurable quantity 
if one works at high $q$ where
$S({\mathbf q},E)\sim S_{\rm IA}({\mathbf q},E)$.
In a uniform Bose-Einstein gas $\rho^{(1)}(s_x,0,0)$ tends to 
a constant when $s_x$ is large. 
In a finite system $\rho^{(1)}(s_x,0,0)$ always
tends to zero when $s_x\to\infty$.
The typical length over which $\rho^{(1)}$ decreases 
can be of the order of the size of the sample or smaller depending
on the degree of coherence.
Using, for example, the Gaussian profile (\ref{exp}) for $S({\mathbf q},E)$
one finds
\begin{equation}
\rho^{(1)}(s_x,0,0)\sim N\exp{\left[-\frac{s_x^2}{2\chi^2_x}\right]}\;,
\label{cohe2}
\end{equation}
with $\chi_x=R_{\perp}\sqrt{3/8}$. One can see from 
Eq. (\ref{cohe2}) that $\chi_x$ plays the role of a  coherence length
\cite{notacohe}, which turns out to be of the order of the size
of the system. This result reflects the fact that in a
Bose-Einstein condensate the Heisenberg inequality
$\Delta R\Delta p_c\ge\hslash/2$ is close to an identity.
\end{subsection}

\begin{subsection}{Eikonal Expansion}
In order to describe the transition between the LDA and IA regimes
discussed above and to better understand the corresponding
conditions of applicability, it is useful to evaluate the high
energy solutions of the Bogoliubov equations (\ref{Bogu}-\ref{Bogv}) 
using an eikonal expansion \cite{Landauqm,notaeiko}. 
In the large $q$ limit, where 
we are interested in the solutions with energy $\epsilon$
much larger than the chemical potential $\mu$, one can neglect 
the function $v({\mathbf r})$ in Eq.(\ref{Bogu}) and look for a 
solution of the form $u({\mathbf r})=
\exp{[i{\mathbf p}_f\cdot{\mathbf r}/\hslash]}\tilde{u}({\mathbf r})$,
where ${\mathbf p}_f$ is the momentum of the excitation and
$\tilde{u}({\mathbf r})$ is a slowly varying function. 
Keeping only terms with first spatial derivative of
$\tilde{u}({\mathbf r})$ (eikonal approximation), 
the solution of (\ref{Bogu}) with 
energy $\epsilon =p^2_f/2m$ takes the form:
\begin{equation}
u({\mathbf r})\simeq\exp{\left[i\frac{{\mathbf p}_f\cdot{\mathbf r}}
{\hslash}\right]}
\exp{\left[-i\frac{m}{p_f\hslash}\int_0^x\!dx^{\prime}V_{\rm eff}
(x^{\prime},y,z)\right]}\;,
\label{uf}
\end{equation}
where the effective potential $V_{\rm eff}({\mathbf r})$, 
calculated in the Thomas Fermi limit, is equal to $gn({\mathbf r})$ 
inside and to $V_{\rm ext}({\mathbf r})-\mu$ outside the condensate. 
At high $q$ the main contribution to the dynamic structure factor 
(\ref{sqobog}) arises from the excited states with 
${\mathbf p}_f\sim{\mathbf q}$. 
This has been taken into account in the eikonal correction 
$\tilde{u}({\mathbf r})$ (second factor of Eq. (\ref{uf})), where
${\mathbf p}_f$ was chosen along the $x$-axis, i.e. the axis fixed by 
the vector ${\mathbf q}$.
Notice that in the eikonal approximation the free particle solution 
$e^{i{\mathbf p}_f\cdot{\mathbf r}/\hslash}$ is modified by the 
interactions only through a change of the phase.
The importance of such a correction in the behaviour of the
dynamic structure factor depends on the
maximum phase deviation of $u({\mathbf r})$ from a pure plane wave,
which is determined by the Born parameter $b$
\begin{equation}
b=\frac{\mu}{E_{\rm r}}\frac{qR_{\perp}}{\hslash}\;,
\label{Born}
\end{equation}
where $E_{\rm r}=q^2/2m$ and $R$ is the Thomas-Fermi radius (\ref{RTF}). 
Differently from the ratio $\mu/E_{\rm r}$, the Born parameter depends 
explicitly on the size of the atomic cloud.

From Eq. (\ref{Bogv}) one gets, in first approximation, 
\begin{equation}
v({\mathbf r})\simeq-\frac{mgn({\mathbf r})}{p^2_f}
u({\mathbf r})\;.
\label{vf}
\end{equation}
Notice that inclusion of $v({\mathbf r})$ in  Eq. (\ref{Bogu}) 
for $u({\mathbf r})$ would result in a higher order correction.
Making the transformation ${\mathbf p}_f={\mathbf p}+{\mathbf q}$, 
the dynamic structure factor (\ref{sqobog}) takes the form:
\begin{equation}
S({\mathbf q},E)=\int\!d{\mathbf p}|M|^2\delta
\left(E-\frac{q^2}{2m}-\frac{qp_x}{m}\right)\;,
\label{sqoeiko}
\end{equation}
with
\begin{equation}
M=\frac{1}{(2\pi\hslash)^{3/2}}
\int\!d{\mathbf r}\psi_0({\mathbf r})\exp{\left[i\frac{m}{\hslash q}
\int_0^x\!dx^{\prime}V_{\rm eff}(x^{\prime},y,z)\right]}
\left(1-\frac{gn({\mathbf r})}{q^2/m}\right)
\exp{\left[-i\left(\frac{{\mathbf p}\cdot{\mathbf r}}{\hslash}\right)
\right]}\;,
\label{M}
\end{equation}
where we have approximated $p_f\sim q$ in the evaluation of 
$v({\mathbf r})$ and in the eikonal correction and neglected the
kinetic energy term $p^2/2m$ in the argument of the $\delta$-function.
This is a very accurate approximation if one works with large samples,
where the Thomas-Fermi approximation 
$\psi_0({\mathbf r})=\sqrt{\big(\mu-V_{\rm ext}({\mathbf r})\big)/g}$ 
applies.

If the Born parameter (\ref{Born}) is small, then the eikonal 
correction can be neglected and, ignoring the small term 
$gn({\mathbf r})m/q^2$ in Eq. (\ref{M}), one recovers the IA 
result (\ref{sqoia}).
Conversely, if $b$ is large one finds a different behaviour. 
In this case the main contribution to the double integral $|M|^2$ 
arises from the region where $|x_1-x_2|\sim R_{\perp}/b\ll R_{\perp}$ and 
the eikonal correction to the relative phase 
$(m/\hslash q)\int_{x_2}^{x_1}\!dx^{\prime}V_{\rm eff}
(x^{\prime},y,z)$  can be consequently written as 
$(m/\hslash q)V_{\rm eff}(x,y,z)(x_1-x_2)$, 
with $x=(x_1+x_2)/2$. 
In these expressions we have set $y_1=y_2=y$ and $z_1=z_2=z$ as a
consequence of the integration on $p_y$ and $p_z$ in
Eq. (\ref{sqoeiko}).
By integrating with respect to the relative variable
$x_1-x_2$ and to $p_x$ one finally recovers the LDA result (\ref{sqold}) 
for large $q$, where 
$\epsilon({\mathbf q},{\mathbf r})\simeq q^2/2m+gn({\mathbf r})$.
In conclusion the eikonal approximation (\ref{sqoeiko}-\ref{M}) provides the
proper description of the dynamic structure factor in the
Thomas-Fermi limit in all the regimes of momentum transfer ranging 
from the LDA to the IA, provided $\mu\ll E_{\rm r}$.
It is interesting to notice that the Born parameter (\ref{Born}) fixes
also the ratio between the widths (\ref{Deltahighq}) and (\ref{Gwidth2})
of the dynamic structure factor calculated in the IA and LDA limits
respectively. In fact one has
\begin{equation}
\frac{\Delta_{\rm LDA}}{\Delta_{\rm IA}}=\frac{b}{14}\;,
\label{wratio}
\end{equation}
so that the comparison between the two widths provides an
equivalent criterion for the applicability of the
two opposite approximations. 
The transition between the LDA and the IA takes place when the
ratio (\ref{wratio}) is close to unity.
Using the relation $\mu =\hslash^2/2m\xi^2$, where $\xi$ is the
healing length (\ref{healing}) calculated at the center of the trap,
this corresponds to the value $q=\hslash R_{\perp}/14\xi^2$ for the
momentum transfer. Notice that this value is much
larger than the inverse of $\xi$, since in the
Thomas-Fermi regime $R_{\perp}\gg\xi$ (see also Fig. 11). 
For example, in the cloud of sodium
atoms explored in Ref. \cite{Bragg1} with peak density $n(0)=3.8\,
10^{14}$ cm$^{-3}$, corresponding to $b\simeq14$, one has
$1/R_{\perp}\simeq 0.1$ $\mu$m$^{-1}$, $1/\xi\simeq 5$ $\mu$m$^{-1}$,
and $R_{\perp}/14\xi^2\simeq 20$ $\mu$m$^{-1}$.

Eqs. (\ref{sqoeiko}-\ref{M}) can be easily calculated numerically in
all the regimes between the LDA and the IA. The comparison with the available
experimental results (see Fig. \ref{fig5}) is rather good and  explains the
deviation of the observed signal from the LDA as
well from the IA predictions. The width of the dynamic structure
factor in general is well reproduced by the quadrature expression
$\sqrt{\Delta_{\rm LDA}^2+\Delta_{\rm IA}^2}$ accounting for both
the LDA and the IA widths (see also \cite{Danthesis}).
\end{subsection}
\end{section}

\begin{section}{Sum rules}
The conditions of applicability of the local density and
impulse approximations presented in the previous section can also be
discussed using a sum rule approach \cite{Lipparini}
which allows for an exact determination of the width of the
dynamic structure factor in some relevant limiting cases.

We have already introduced in the previous section
the moments of the dynamic structure factor
relative to the operator $\rho_{{\mathbf q}}$.
In terms of the matrix elements $\langle n|\rho_{{\mathbf q}}|0\rangle $
of the density operator one can write
\begin{equation}
m_k({\mathbf q}) =\sum_{n\neq 0}
|\langle n|\rho_{{\mathbf q}}|0\rangle |^2(E_n-E_0)^k=
\int_{0^{+}}^{\infty}\!dES({\mathbf q},E)E^k\;,
\label{moments2}
\end{equation}
and, using the closure relationship $\sum_n|n\rangle\langle n|=\textbf{1}$,
one can easily express the moments $m_k({\mathbf q})$ in terms of mean
values on the ground state $|0\rangle$ and of commutators between the
Hamiltonian and the operator $\rho_{{\mathbf q}}$. 
Using the property
$S({\mathbf q},E)=S(-{\mathbf q},E)$, holding in the presence of
parity or time-reversal invariance, 
we find the following results for the lowest moments:
\begin{align}
m_0(q)&=
\langle\rho_{{{\mathbf q}}}^{\dagger}\rho_{{{\mathbf q}}}\rangle
-|\langle\rho_{{\mathbf q}}\rangle|^2=NS(q)
\label{m0}\\
m_1(q)&=\frac{1}{2}
\langle[\rho_{{{\mathbf q}}}^{\dagger},[H,\rho_{{{\mathbf q}}}]]\rangle =
N\frac{q^2}{2m}
\label{m1}\\
m_2(q)&=
\langle[\rho_{{{\mathbf q}}}^{\dagger},H][H,\rho_{{{\mathbf q}}}]\rangle =
N\bigg[[2-S(q)]\bigg[\frac{q^2}{2m}\bigg]^2+\frac{\hslash^2q^2}{m^2}
D^x(q)\bigg]
\label{m2}\\
m_3(q)&=\frac{1}{2}
\langle[[\rho_{{{\mathbf q}}}^{\dagger},H],[H,[H,\rho_{{{\mathbf q}}}]]]
\rangle = \nonumber\\
&=N\bigg[\bigg[\frac{q^2}{2m}\bigg]^3+
4\bigg[\frac{q^2}{2m}\bigg]^2\bigg(3\frac{E_{\rm kin}^x}{N}+
\frac{E_{\rm int}}{N}\bigg)
+\frac{q^2}{2m}\frac{\hslash^2}{m}
\langle\partial^2_xV_{{\rm ext}}\rangle\bigg]\;,
\label{m3}
\end{align}
where we took the vector ${\mathbf q}$ along the $x$ axis.
Notice that in Eq. (\ref{m0}) we have subtracted the elastic
contribution $|\langle\rho_{{\mathbf q}}\rangle|^2$.
The kinetic structure function $D^x(q)$ is defined by
\begin{equation}
D^x(q)=\frac{1}{N}\int\!d{\mathbf r}_1d{\mathbf r}_2\cos{[q
(x_1-x_2)]}\mbox{\boldmath $\nabla$}_1^x
\mbox{\boldmath $\nabla$}_2^x\rho^{(2)}({\mathbf r}_1,{\mathbf r}_2;
{\mathbf r}_1^{\prime},{\mathbf r}_2^{\prime})|_{{\mathbf r}_1=
{\mathbf r}_1^{\prime},{\mathbf r}_2={\mathbf r}_2^{\prime}}\;,
\label{Dq}
\end{equation}
where 
\begin{equation}
\rho^{(2)}
({\mathbf r}_1,{\mathbf r}_2;{\mathbf r}_1^{\prime},{\mathbf r}_2^{\prime})=
N(N-1)\int\!d{\mathbf r}_3d{\mathbf r}_4...d{\mathbf r}_N
\psi^*({\mathbf r}_1,{\mathbf r}_2,...,{\mathbf r}_N)
\psi ({\mathbf r}_1^{\prime},{\mathbf
  r}_2^{\prime},{\mathbf r}_3,{\mathbf r}_4,...,{\mathbf r}_N)
\label{rho2}
\end{equation} 
is the two-body density matrix.
In the asymptotic limit $q\to\infty$ this function is related to the
kinetic energy of the system \cite{Feenberg}:
\begin{equation}
\lim_{q\to\infty}D^x(q)=\frac{2m}{\hslash^2}\frac{E_{{\rm kin}}^x}{N}\;.
\label{Dqekin}
\end{equation}

The $m_3$ sum rule (\ref{m3}) has been obtained evaluating
the commutators with the effective Hamiltonian
\begin{equation}
H=\sum_i\left[\frac{p_i^2}{2m}+V_{{\rm ext}}({\mathbf r}_i)\right]
+g\sum_{i<j}\delta ({\mathbf r}_i-{\mathbf r}_j)\;,
\label{Hmf}
\end{equation}
and using the corresponding ground state of Gross-Pitaevskii
theory. 
In particular $E_{\rm int}=g\int\!d{\mathbf r}n({\mathbf r})^2/2$
corresponds to the expectation value of the two body interaction energy.

The f-sum rule (\ref{m1}) is model independent \cite{Pines} and is
satisfied by both the LDA and IA as can be explicitly shown by integrating
the corresponding dynamic structure factors.
The other sum rules are instead correctly reproduced only in
suitable ranges of momenta, which thereby provide the corresponding
regimes of applicability of the two approximations.

The non energy weighted sum rule (\ref{m0}) coincides with the
inelastic static structure
factor $S({\mathbf q})$, a quantity of high interest, directly
related to the Fourier transform of the diagonal two-body density matrix:
\begin{equation}
S({\mathbf q})=1+\frac{1}{N}\int\!d{\mathbf r}_1d{\mathbf r}_2\;
e^{i{\mathbf q}\cdot({\mathbf r}_1-{\mathbf r}_2)}
\big[\rho^{(2)}({\mathbf r}_1,{\mathbf r}_2;
{\mathbf r}_1,{\mathbf r}_2)-
n({\mathbf r}_1)n({\mathbf r}_2)\big]\;.
\label{sqdef}
\end{equation}
The LDA prediction for $S({\mathbf q})$ has been already discussed in the
previous section [see Eq.(\ref{Sqdan})]. This prediction is
expected to hold in all regimes of small and high $q$ except, of
course, when $qR < \hslash$. In fact the static structure factor
$S({\mathbf q})$ is not sensitive to the Doppler broadening which instead
affects other moments of the dynamic structure factor. An
interesting property of the static structure factor is the
occurrence of a $1/q^2$ correction to the large $q$ asymptotic
value:
\begin{equation}
S(q)=1-\frac{8}{7}\frac{m\mu}{q^2}\;.
\label{sqcorrected}
\end{equation}
Such a correction is a peculiarity of dilute Bose gases \cite{Huang}
and is directly
related to the shift of the average excitation energy 
given by the Feynman ratio $m_1/m_0 = E_{{\rm r}}+4\mu/7$.
The IA does not instead predict any $q$-dependence for the static form
factor and consequently fails in reproducing the shift of the peak.

The $m_2$ sum rule is also interesting for understanding the difference
between the LDA and the IA. For large $q$ this sum rule contains two
corrections to the leading asymptotic value:
\begin{equation}
\lim_{q\to\infty} m_2(q)
= \bigg[\frac{q^2}{2m}\bigg]^2+\frac{q^2}{2m}
\bigg[\frac{4}{7}\mu+4 \frac{E_{{\rm kin}}^x}{N}\bigg]\;.
\label{m2largeq}
\end{equation}
The first correction, fixed by the chemical potential,
arises again from the large q behaviour (\ref{sqcorrected}) of $S(q)$
which enters the expression (\ref{m2}) for $m_2$.
The second contribution, proportional to the kinetic energy, arises
form the  kinetic structure factor $D^x(q)$.
The first correction is correctly given by the LDA, the latter by the IA.
It is worth noticing that if one
calculates the rms width (\ref{width}) only the kinetic energy
term survives in the large $q$ limit.
This  confirms the correctness of the impulse approximation in reproducing
the width of the dynamic structure factor at high $q$.

Finally the $m_3$ sum rule is interesting because it can be
explicitly evaluated for  any value of $q$. For a uniform gas
($V_{{\rm ext}}=0$ and $E_{{\rm kin}}=0$) the ratio
$\sqrt{m_3/m_1}$ coincides with the Bogoliubov excitation spectrum
(\ref{Bogspectrum}). In the presence of harmonic trapping it is
instructive to calculate Eq. (\ref{m3}) in the small momentum
transfer limit, where only the last term, containing the external
potential, survives and the ratio $\sqrt{m_3/m_1}$ coincides with
the frequency $\omega_{\perp}$ of the dipole mode.
Indeed, the dipole mode is the only mode excited by
the density operator $\rho_{{\mathbf q}}=\sum_j
e^{i{\mathbf q}\cdot{\mathbf r}_j/\hslash}$ in the $q\to 0$ limit.

Before concluding this section we stress  that the 
results discussed above hold in the Bogoliubov regime
$qa\ll\hslash$. For larger momenta, relevant for example in the case
of superfluid helium, a different behaviour takes place. 
For example, result (\ref{sqcorrected}) for the static structure factor 
$S(q)$ is no longer valid. Particular attention should be
also paid to the kinetic energy which characterizes the large
$q$ behaviour of $m_2$  and hence of the rms width. 
One should in fact distinguish between the kinetic energy of the
condensate and the full kinetic energy of the system which, even
in dilute Bose gases, is dominated at $T=0$ by the
high momentum components of $n({\mathbf p})$. 
The former is given by (\ref{EkinTF})
and becomes smaller and smaller as $R$ increases. 
The latter is instead of the order of the interaction energy and 
would determine the rms width 
of the dynamic structure factor for momentum transfer larger than $\hslash /a$.
\end{section}

\begin{section}{Scaling and impulse approximation}
In the previous sections we have often referred to the impulse
approximation as the proper theory to describe the high
$q$ response of the system. At the same time we have pointed out
that the IA does not account for the mean field shift
(\ref{shift}) of the peak energy occurring at high $q$, which is
instead correctly predicted by the LDA. In this section we discuss
in what sense the IA provides the exact asymptotic description of
the dynamic structure factor. The discussion is simplified by
using scaling, a concept already employed in other many body
systems, including atomic nuclei \cite{Sick}, liquids and solids
\cite{Sears,Sokol}. Let us introduce the scaling variable
\begin{equation}
Y=\frac{m}{q}\left(E-\frac{q^2}{2m}\right)\;,
\label{Yvar}
\end{equation}
which is the relevant variable to describe the asymptotic behaviour of
the dynamic structure factor.
We define the scaling function $F_0(Y)$ according to the asymptotic
behaviour
\begin{equation}
F_0(Y)=\lim_{q\to\infty}\frac{q}{m}\frac{S({\mathbf q},E)}{N}\;,
\label{scaling1}
\end{equation}
where, in the limit, the excitation energy $E$ varies with $q$ in order
to keep the value of $Y$ fixed.
Comparison with Eq. (\ref{sqoia2}) shows that
$F_0(Y)$ coincides with the longitudinal momentum distribution
\begin{equation}
F_0(Y)=\frac{1}{N}\int\!dp_ydp_zn(Y,p_y,p_z)\;.
\label{scaling2}
\end{equation}
In terms of the scaling function $F_0(Y)$ the rms width takes the form
\begin{equation}
\Delta_{\rm rms}=\frac{q}{m}\sqrt{\int_{-\infty}^{\infty}\!dYF_0(Y)Y^2}\;.
\label{widthscaling}
\end{equation}
Furthermore one has $\int_{-\infty}^{\infty}\!dYF_0(Y)=1$, and
$\int_{-\infty}^{\infty}\!dYYF_0(Y)=0$.

The scaling result (\ref{scaling1}-\ref{scaling2})
hold for a wide class of many body systems interacting with
realistic two-body potentials, and is not restricted to dilute gases
(see for example \cite{Carraro} and references therein).
This means that corrections to the IA, due to final state interactions,
give a vanishing contribution to the limit (\ref{scaling1}).
In general the following expansion holds:
\begin{equation}
\frac{q}{m}\frac{S({\mathbf q},E)}{N}=F_0(Y)+
\frac{m}{q}F_1(Y)+\left(\frac{m}{q}\right)^2F_2(Y)+\dots\;.
\label{espansione}
\end{equation}
In the presence of parity or time
reversal symmetry the scaling function $F_0(Y)$ is symmetric:
$F_0(Y)=F_0(-Y)$. Conversely one has: $F_1(Y)=-F_1(-Y)$. This
suggests that the proper symmetrization
$[S({\mathbf q},E)+S({\mathbf q},2E-E_{\rm r})]/2$ of the measured
signal with respect to the recoil energy $E_{\rm r}=q^2/2m$, would
ensure a faster convergence to the scaling limit \cite{Sears}.
From an experimental point of view the direct verification of
scaling, by changing $q$ and $E$ keeping $Y$ fixed, is likely the
safest criterion for checking the achievement of
the IA regime. A peculiarity of dilute gases is that it is
possible to reach the scaling regime for values of momenta where
the Bogoliubov theory is still applicable.

For a trapped Bose gas, the eikonal Eqs. (\ref{sqoeiko}-\ref{M}) can
be easily expanded for small values of the Born
parameter (\ref{Born}), corresponding to high values of $q$ ($b=2m\mu
R_{\perp}/\hslash q$). One finds: 
\begin{equation}
\frac{q}{m}\frac{S({\mathbf q},E)}{N}=
\frac{R_{\perp}}{\hslash}\big[f_0(y)+bf_1(y)+\cdots\big]\;,
\label{reducedexp}
\end{equation}
where $f_0(y)$ and $f_1(y)$ are dimensionless functions of the 
variable $y=(R_{\perp}/\hslash )Y$, directly related to the scaling functions
of Eq. (\ref{espansione}): 
$F_0(Y)=R_{\perp}f_0(y)/\hslash$ and 
$F_1(Y)=qR_{\perp}bf_1(y)/m\hslash$.
The function $f_0(y)$ is given by:
\begin{equation}
f_0(y)=\frac{15}{16}\int
\!d\tilde{p}_yd\tilde{p}_z\left[\frac{
J_2\big(\sqrt{y^2+\tilde{p}_y^2+\tilde{p}_z^2}\big)}
{y^2+\tilde{p}_y^2+\tilde{p}_z^2}\right]^2\;,
\label{fTF}
\end{equation}
and is shown in Fig. \ref{fig6} together with the Gaussian expansion
\begin{equation}
f_0(y)\simeq\frac{15\pi}{192}\exp{\left[-\frac{3}{16}y^2\right]}\;,
\label{gauss}
\end{equation}
yielding the results (\ref{exp}-\ref{Gwidth2}) for the dynamic
structure factor.
The function $f_1(y)$ is given by
\begin{multline}
f_1(y)=\frac{15}{2}
\int_0^{\infty}\!dp_{\perp}p_{\perp}
\frac{J_2\big(\sqrt{y^2+p_{\perp}^2}\big)}{y^2+p_{\perp}^2}\times\\
\int_0^1\!dr_{\perp}r_{\perp}J_0\big(r_{\perp}p_{\perp}\big)
\int_0^{\sqrt{1-r_{\perp}^2}}\!dx\sqrt{1-x^2-r_{\perp}^2}\sin{(yx)}
x\left(1-\frac{x^2}{3}-r_{\perp}^2\right)\;,
\label{F1}
\end{multline}
and is also shown in Fig. \ref{fig6}.

Starting from Eq. (\ref{reducedexp}) one can evaluate the 
shift of the peak with respect to the recoil energy $E_{\rm r}$
due to the first correction to the IA. 
Imposing the condition $\partial_ES({\mathbf q},E)=0$ one finds, 
after some straightforward algebra, 
$E_{\rm peak}=E_{\rm r}+2\mu/3$, 
showing that result (\ref{epeak}) for the line
shift holds not only in the high $q$ LDA regime where $b$ is large,
but also for small values of $b$.
\end{section}

\begin{section}{Dynamic structure factor and vortices}
The study of vortices in trapped Bose gases is presently a challenging
topics of both theoretical and experimental investigation.
First experimental evidence of a vortices
has been recently reported \cite{vortexJILA,vortexENS}.
On the theoretical side the structure of vortices,
the corresponding stability conditions, as well as their consequences
on the dynamic behaviour of the condensate
have already attracted the attention of many physicists.
The identification of suitable methods of detection has also been the
object of theoretical investigation. These include the
expansion of the condensate \cite{Michele}, the shift of the collective
excitation frequencies \cite{Francesca} and the occurrence of dislocations
in the interference patterns \cite{dislocazione}.

In this section we show that the measurement of the dynamic
structure factor in the IA regime would represent a powerful tool
to reveal vortices in a trapped Bose gas. In fact a 
vortex strongly affects the momentum
distribution of the system. This can be easily understood 
by noting that the kinetic energy of a trapped condensate is roughly
doubled by the addition of a vortex \cite{Lundh}. 
In contrast, the density distribution of a condensate in the
Thomas-Fermi regime is only slightly modified by the vortex whose size
is small compared to that of the gas. 

In the presence of a quantized vortex aligned
along the $z$-axis the wave function of the condensate takes the form
\begin{equation}
\psi({\mathbf r})=e^{i\varphi}\psi_0(r_{\perp},z)\;,
\label{psiv}
\end{equation}
where $\psi_0(r_{\perp},z))$ is the 
solution of the Gross-Pitaevskii equation \cite{LP,GP}
\begin{equation}
\left[-\frac{\hslash^2\nabla^2}{2m}+\frac{\hslash^2}{2mr_{\perp}^2}+
\frac{m}{2}(\omega_{\perp}^2r_{\perp}^2+\omega_z^2z^2)+
g\psi_0(r_{\perp},z)^2\right]\psi_0(r_{\perp},z)=\mu\psi_0(r_{\perp},z)\;,
\label{GPvortex}
\end{equation}
which contains the additional centrifugal term 
$\hslash^2/2mr_{\perp}^2$. Solutions of Eq. (\ref{GPvortex}) have been
obtained numerically in \cite{GPnum}.
The density distribution $|\psi_0(r_{\perp},z)|^2$ exhibits a hole
whose size is of the order of the healing length (\ref{healing}) of the gas
$\xi$, which, in the Thomas-Fermi limit, is much 
smaller than the size of the condensate.
Also in momentum space the distribution exhibits a hole as shown in
Fig. \ref{fig7}. 
This is the consequence of the phase in Eq. (\ref{psiv}), which gives a
vanishing value to the integral (\ref{Fourier}) at $p_{\perp}=0$,
where $p_{\perp}$ is the radial component of the momentum vector
${\mathbf p}$. The size of the hole is of the order of $\hslash/R$, 
and consequently comparable to the total size of the
condensate in momentum space.
This can be easily seen calculating the momentum distribution in
the Thomas-Fermi limit. In this limit the main effect of the vortex on
the momentum distribution arises from the phase $e^{i\varphi}$,
and one can safely use for $\psi_0$ the Thomas-Fermi expression
$\sqrt{(\mu-V_{\rm ext}({\mathbf r}))/g}$, holding in the absence of
the vortex. The result for $n({\mathbf p})$ can then be written in the form
\begin{equation}
n_{\rm TF}({\mathbf p})=N\frac{R_{\perp}^3}{\hslash}
\frac{15}{16}\frac{\lambda}{\pi^4}
\left|
\int_0^{2\pi}\!\!d\varphi\,e^{i\varphi}\int_0^{\infty}\!\!dr_{\perp}r_{\perp}
\int_0^{\sqrt{1-r_{\perp}^2}}\!\!dz\,
e^{i\tilde{{\mathbf p}}\cdot{\mathbf r}}
\sqrt{1-r_{\perp}^2-z^2}\right|^2\;,
\label{vortexTF}
\end{equation}
where $\tilde{{\mathbf p}}$ is the scaled momentum vector
$\tilde{{\mathbf p}}\equiv(p_x,p_y,p_z/\lambda)R_{\perp}/\hslash$, already
introduced in sec. III B.
Notice that in the Thomas-Fermi limit the effect of the vortex is 
factorized through a dimensionless integral.

In Fig. \ref{fig8} we report the dynamic structure factor
calculated in the IA (see Eq. (\ref{sqoia})) with and without the
vortex. The calculation was carried out for a gas of $N=10^4$
$^{87}$Rb atoms trapped in a disk type geometry ($\lambda=\sqrt{8}$). 
For this low density sample the IA is very accurate.
The double peak structure in $S({\mathbf q},E)$ reflects the occurrence of a
peculiar Doppler effect, and represents a clear signature of the
vortex. In fact the vortex generates a velocity field in the
condensate with significant components both parallel and
antiparallel to the momentum transfer ${\mathbf q}$. 
\end{section}

\begin{section}{Interference effects in momentum space}
Finally, let us discuss the dynamic structure factor in terms of 
the occurrence of interference phenomena in momentum space. 
Interference has been so far
investigated in coordinate space by imaging two overlapping Bose
Einstein condensates \cite{MIT}. 
However, even if the two condensates do not
overlap in space, they can interfere in momentum space
\cite{Lev}. This opens the possibility of investigating
interference while avoiding any interaction between the two condensates.

Consider, for example, a double well potential and let us assume, for
simplicity, that the condensates in the two wells (condensates $a$ and
$b$ respectively) have the same number of atoms (see Fig. \ref{fig9}). 
If the distance $d$ between the two wells is large enough to avoid 
overlapping, and if the potential acting on the condensates $a$ and
$b$ can be obtained by a simple translation, then their wave functions
can be written as:
\begin{align}
\psi_a({\mathbf r})&=\psi_0\left({\mathbf r}-\frac{\textbf{d}}{2}\right)
\label{psia}\\
\psi_b({\mathbf r})&=\psi_0\left({\mathbf r}+\frac{\textbf{d}}{2}\right)\;,
\label{psib}
\end{align}
where $\psi_0$ is the solution of the Gross Pitaevskii equation for
each condensate.
The Fourier transforms of Eqs. (\ref{psia}-\ref{psib}) hence read:
\begin{eqnarray}
\phi_a({\mathbf p})&=&e^{ip_xd/2\hslash}\phi_0({\mathbf p})
\label{phia}\\
\phi_b({\mathbf p})&=&e^{-ip_xd/2\hslash}\phi_0({\mathbf p})\;,
\label{phib}
\end{eqnarray}
having taken the displacement between the two wells along the $x$
axis. Under the above conditions any linear combination
\begin{equation}
\psi_{\rm C}({\mathbf r})=\psi_a({\mathbf
  r})+e^{i\varphi}\psi_b({\mathbf r})
\label{comblin}
\end{equation}
of the wavefunctions (\ref{psia}-\ref{psib}) corresponds to a solution
of the Gross Pitaevskii equation. These combinations represent
coherent configurations which exhibit interference patterns in the
momentum distribution:
\begin{equation}
n({\mathbf p})=2n_0({\mathbf p})\left[1+
\cos{\left(\frac{p_xd}{\hslash}+\varphi\right)}\right]\;,
\label{npint}
\end{equation}
where $n_0({\mathbf p})=|\phi_0({\mathbf p})|^2$.
These patterns have interesting consequences on the shape of the
dynamic structure factor which, in the IA, takes the form 
\begin{equation}
S({\mathbf q},E)=2S_0({\mathbf q},E)
\left[1+\cos{\left(\frac{Yd}{\hslash}+\varphi\right)}\right]\;,
\label{sqoiaint}
\end{equation}
where $Y$ is the scaling variable (\ref{Yvar}). The dynamic structure factor
(\ref{sqoiaint}) exhibits fringes with frequency period
\begin{equation}
\Delta\nu=\frac{\Delta E}{h}=\frac{q}{md}\;.
\label{fringes}
\end{equation}
In Fig. \ref{fig10} we show a typical result for $S({\mathbf q},E)$
corresponding to a distance between the two condensates four times 
larger than their radial width.
The position of the fringes depends crucially on the value of
the relative phase between the two condensates. 
\end{section}

\begin{section}{Conclusions}
In this paper we have provided a theoretical discussion of the dynamic 
structure factor $S({\mathbf q},E)$ of a trapped Bose-Einstein
condensate at low temperature. 
A first important aim was the development of the proper many-body
formalism, based on Bogoliubov theory, to describe in a quantitative
way the several interesting features exhibited by $S({\mathbf q},E)$. 
These mainly concern the role of two-body interactions which sizably affect
the response of the system in all the relevant regimes of momentum
transfer. The possibility of providing accurate theoretical
predictions for the dynamic structure factor is particularly appealing
in view of the recent experimental data obtained via two-photon
Bragg scattering. The available data are in general agreement with
theory, and thereby provide a further important proof of the crucial
role played by two-body interactions in these trapped
Bose-Einstein condensed gases. 
Interactions affect the shape of  $S({\mathbf q},E)$ at both small momentum
transfer, where they are responsible for the propagation of phonons, 
and at high momentum where they show up in the shift of the peak with
respect to the free recoil energy $q^2/2m$ as well as in the width
which is sensitive to both mean field and Doppler effects. Various 
approximate schemes have been considered in order to better discuss
the main physical features. 
These schemes are summarized in Fig. 11 where the importance of the 
characteristic length scales of the problem emerges clearly. 
At momentum transfer smaller than the inverse of the size of the
system the response is characterized by the discretized normal
modes of the system.
This regime has been already discussed in other works and has not been 
considered here. At higher momenta the system behaves locally as a
uniform gas. This is the range of applicability of the local density 
approximation (LDA) which successfully describes the excitation of
phonons as well as, at momenta larger than the inverse of the healing
length, the corrections to the free particle motion due to mean field 
interactions. At even higher momenta the response of the system cannot
be longer described locally, because it is sensitive to the momentum 
distribution of the condensate, a quantity associated with long-range 
coherence effects. This regime is well described by the impulse 
approximation (IA), a theory currently employed to investigate the 
quasi-free response of various many-body systems. In our work we have 
limited the use of the IA to momenta smaller than the
inverse of the scattering length, i.e. to the range of applicability
of Bogoliubov theory. In this regime only the momentum distribution of 
the condensate is relevant and one can safely ignore the higher
momentum components of $n({\mathbf p})$ which would be crucial to
describe the response at momentum transfers larger than $\hslash/a$. 
An interesting feature emerging from our analysis is that the
transition between the LDA and the IA regimes is characterized by an  
important physical parameter, the so-called Born parameter
(\ref{Born}) which depends explicitly on the size of the system. 
The corresponding transition takes place at momenta fixed by the 
combination $\hslash R/\xi^2$. We have shown that the
transition can be accounted for by an eikonal treatment of the
solutions of the Bogoliubov equations and a full calculation of 
$S({\mathbf q},E)$ has been presented in this regime, showing  good 
agreement with experiments. 
Many of the relevant features exhibited by the dynamic structure
factor in these trapped Bose gases have been also presented and
discussed using the formalism of sum rules and the concept of scaling.

A second important point emerging from our analysis was that the 
dynamic structure factor in the IA regime of high 
momentum transfer offers a new important investigation tool.   
Actually Bose-Einstein condensation in momentum space provides a
deeper understanding of long-range coherence phenomena in comparison with 
the studies of the density profiles which have been so far  the main
object of investigation.  In our paper we have discussed some
significant examples of such opportunities, including the case of 
quantized vortices and of interference effects occurring in momentum
space. In both cases we have predicted non-trivial features which
should be visible in the dynamic structure factor as a consequence of 
the peculiar behaviour of the order parameter in momentum space. 
In the case of vortices the dynamic structure factor is characterized
by a hole occurring at the recoil energy and whose size is comparable 
to the total width of the signal. 
In a second example we have shown that the dynamic structure 
factor of two spatially separated condensates should exhibit  
interference fringes which are the consequence of the coherence
existing between the two condensates.
\end{section}

\begin{section}{Acknowledgements}
We are very grateful to W.~Ketterle and A.P.~Chikkatur for many
fruitful discussions and for providing experimental data and details.
D.M. S.-K. acknowledges support of Millikan Prize Postdoctoral Fellowship.
This work has been supported by the Istituto Nazionale per la Fisica
della Materia (INFM) through the Advances Research Project on BEC, and by
Ministero dell'Universit\`a e della Ricerca Scientifica e Tecnologica (MURST).
\end{section}

\begin{figure}
\begin{center}
\input{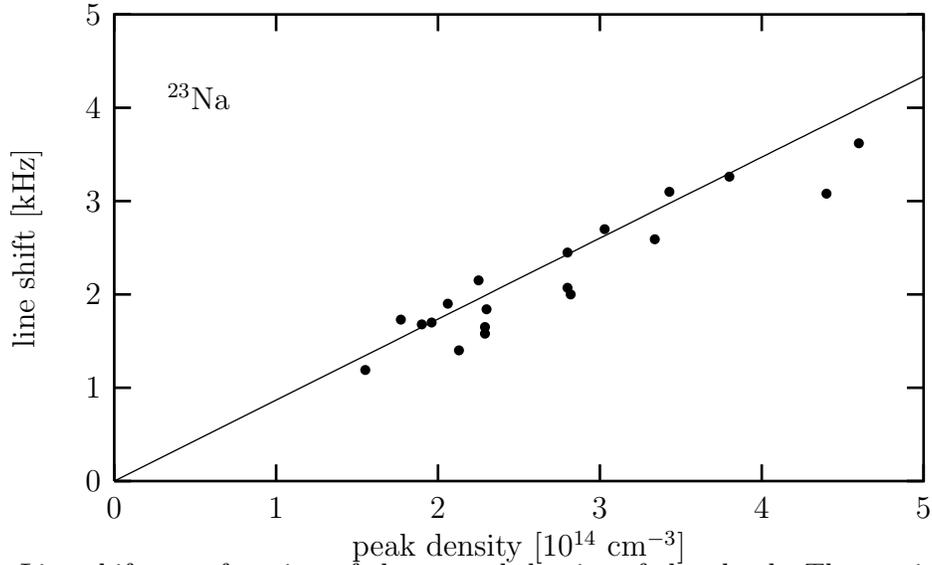}
\caption{Line shift as a function of the
central density of the cloud. The straight line is the theoretical
prediction (\ref{shift}), and the circles are the experimental
points of Ref. \protect\cite{Bragg1}.} 
\label{fig3}
\end{center}
\end{figure}

\begin{figure}
\begin{center}
\input{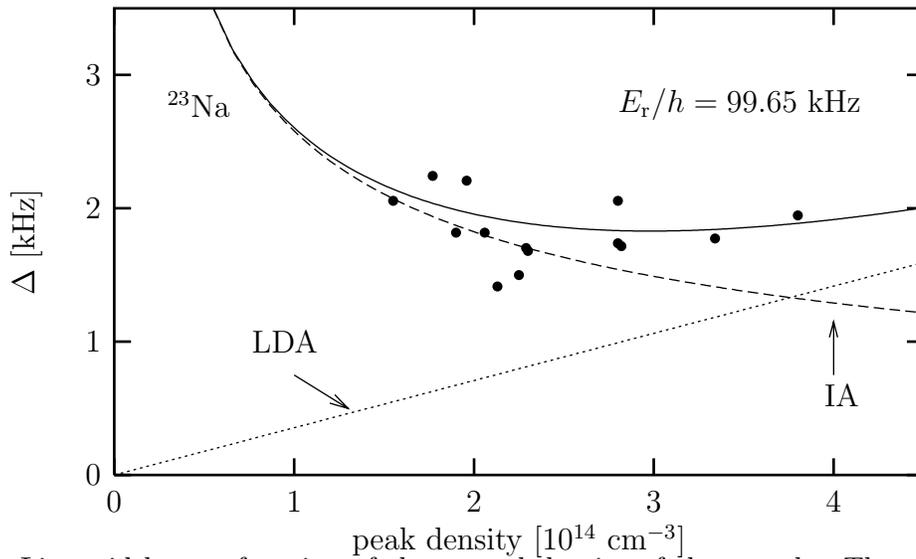}
\caption{Line width as a function of the central density of the sample. 
The solid line is the prediction of the eikonal approximation 
(\ref{sqoeiko}). 
The dashed and dotted lines correspond to the IA and LDA predictions 
(\ref{Gwidth2}) and (\ref{Deltahighq}) respectively. 
The theoretical results are compared with the experimental data of 
\protect\cite{Bragg1}. 
Both the eikonal and the experimental values are obtained
through the Gaussian fit $S(q,\bar{E})\exp{[(E-\bar{E})^2/2\Delta^2]}$ 
to the signal. The momentum transfer is taken along the $x$-axis.
The trapping frequencies are $\omega_{\perp}=2\pi\,195$ Hz and 
$\omega_z=2\pi\,17$ Hz.}
\label{fig4}
\end{center}
\end{figure}

\begin{figure}
\begin{center}
\input{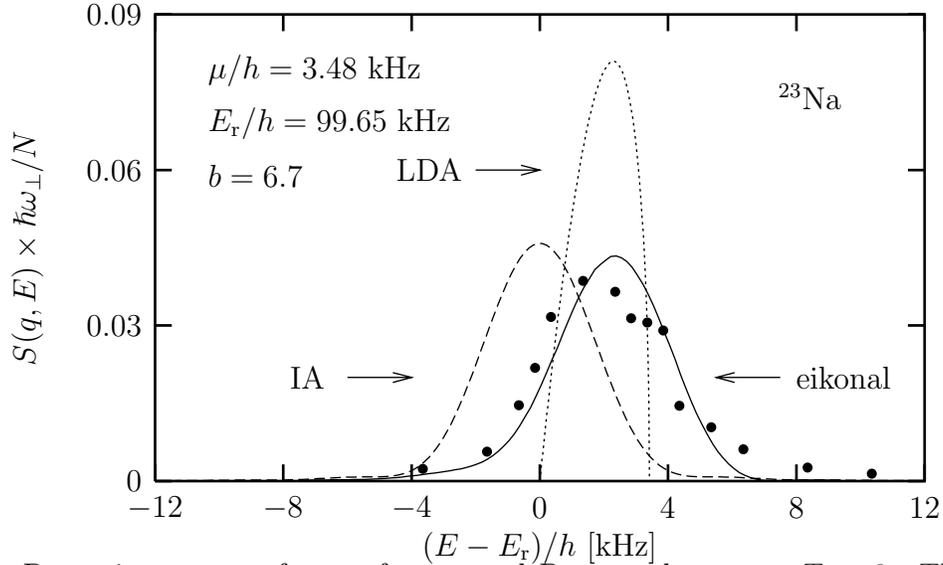}
\caption{Dynamic structure factor of a trapped Bose condensate at $T=0$. 
The numerical predictions of
the eikonal approximation (solid curve), IA (dashed curve) and
LDA (dotted curve) are compared with the experimental data of
Ref. \protect\cite{Bragg1}, normalized to reproduce $S(q)=1$.
The momentum transfer is taken along the $x$-axis.
The trap parameters are the same as in Fig. \ref{fig4}.}
\label{fig5}
\end{center}
\end{figure}

\begin{figure}
\begin{center}
\input{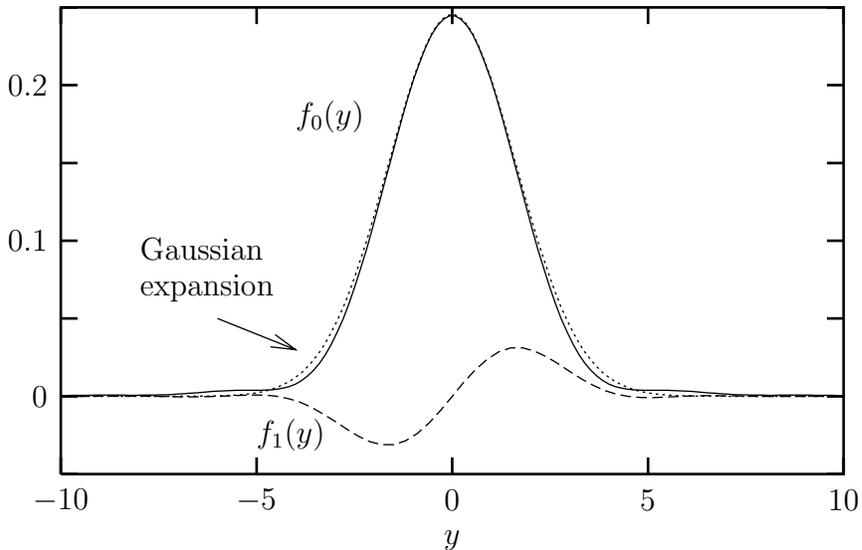}
\caption{Dimensionless scaling functions in Thomas-Fermi regime: 
the $f_0(y)$ function (Eq. (\ref{fTF})) (solid line) is compared with
its Gaussian expansion (\ref{gauss}) (dotted line). The
function $f_1(y)$ (Eq. (\ref{F1})) is also shown (dashed line).}
\label{fig6}
\end{center}
\end{figure}

\begin{figure}
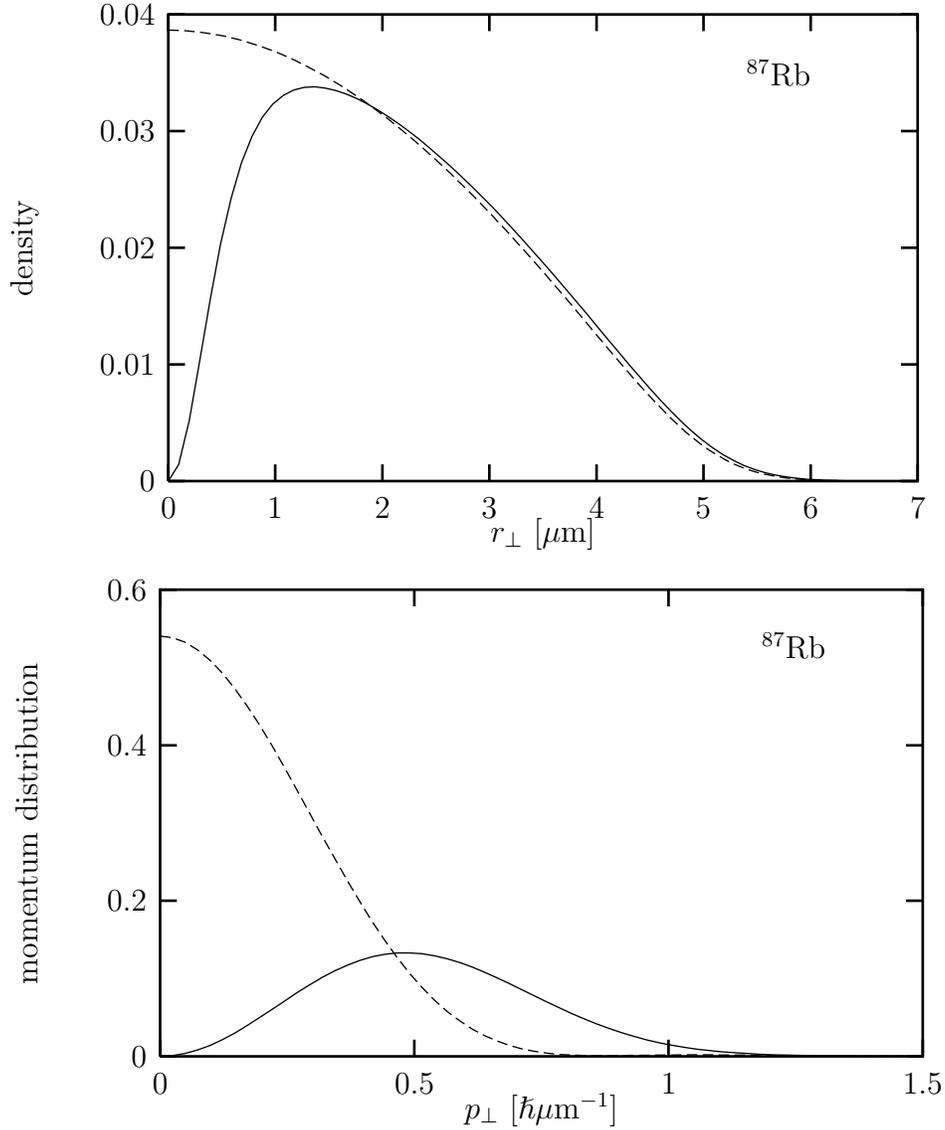

\begin{center}
\input{fig7up.tex}\input{fig7down.tex}
\caption{Density profile (up) and momentum distribution (down) of a 
trapped Bose condensate as a function of the radial variables
$r_{\perp}$ and $p_{\perp}$, integrated along $z$ and $p_z$
respectively, in the absence (dashed line) and in the presence of a 
quantized vortex (full line).
These profiles correspond, in the absence of the vortex, to a central 
density $n(0)\simeq 1\,10^{14}$ cm$^{-3}$. 
The trap parameters are $\omega_z=2\pi\,220$ Hz, and $\lambda =\sqrt{8}$.}
\label{fig7}
\end{center}
\end{figure}

\begin{figure}
\begin{center}
\input{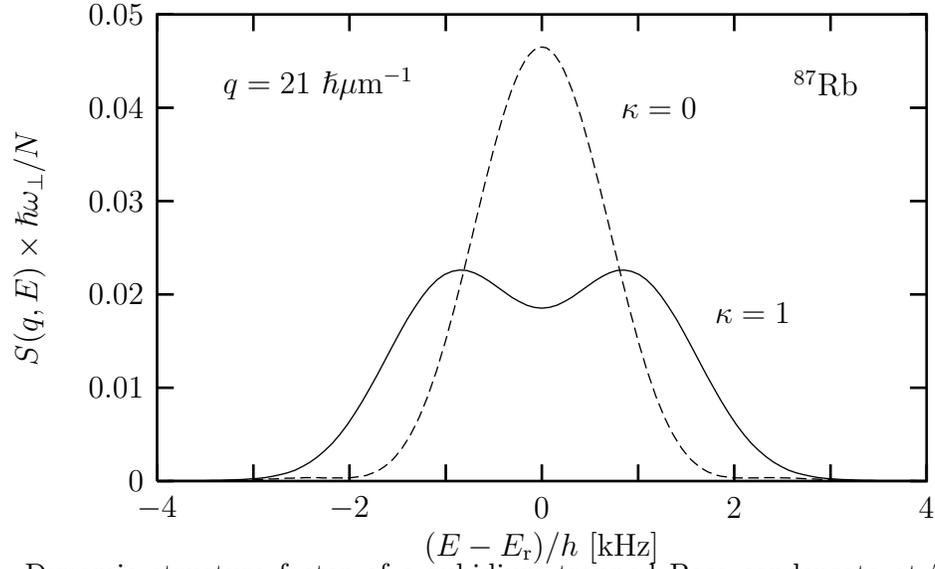}
\caption{Dynamic structure factor of a rubidium trapped Bose
condensate at $T=0$ in the presence (solid line) and in the absence 
(dashed line) of the vortex, calculated using the impulse
approximation (\ref{sqoia}). The momentum transfer is taken along the
$x$-axis.  
The trap parameters are the same as in Fig. \ref{fig7}.}
\label{fig8}
\end{center}
\end{figure}

\newpage
\begin{figure}
\begin{center}
\input{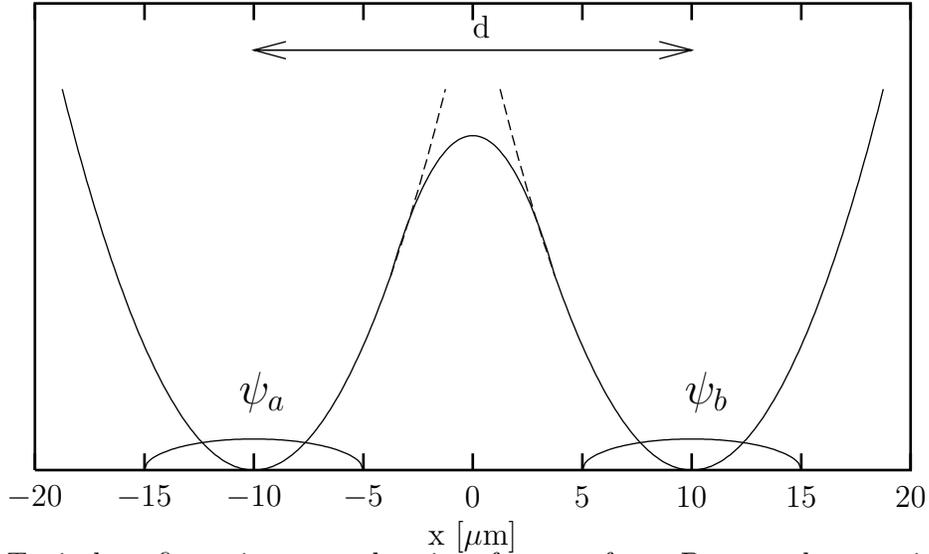}
\caption{Typical configuration to explore interference of two Bose condensates
in momentum space: two harmonic wells are separated 
by a distance $d=20$ $\mu$m; each well confines a condensate of sodium atoms
with radial size $R_{\perp}=5$ $\mu$m and central density 
$n(0)\simeq 0.7\,10^{14}$ cm$^{-3}$.} 
\label{fig9}
\end{center}
\end{figure}

\begin{figure}
\begin{center}
\input{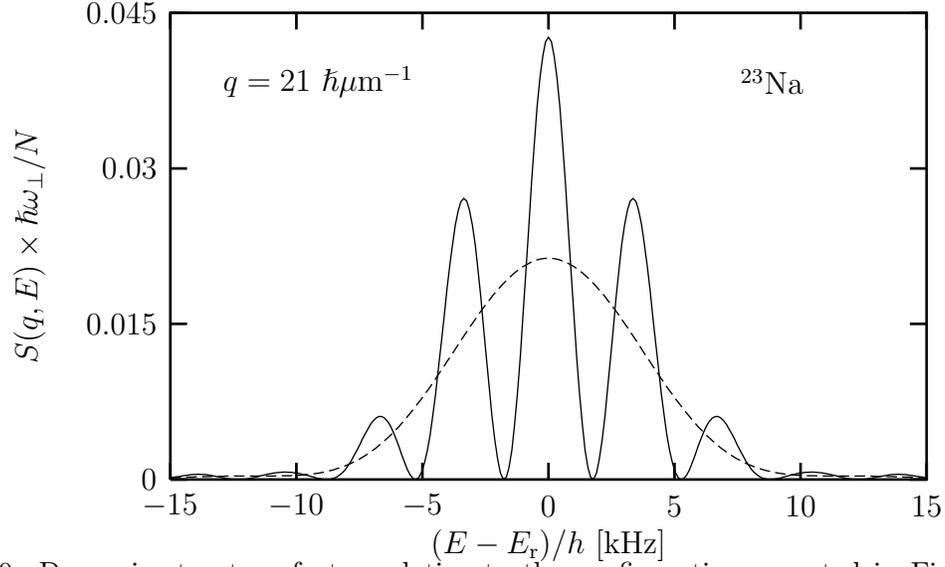}
\caption{Dynamic structure factor relative to the configuration reported in
Fig. \ref{fig9} without (dashed line) and with (solid line) coherence
between the two condensates (Eq. (\ref{sqoiaint})).
The momentum transfer $q$ is taken along the $x$-axis. 
The relative phase between the two condensates is chosen equal to zero.
The trap parameters are the same as in Fig. \ref{fig4}.}
\label{fig10}
\end{center}
\end{figure}

\begin{figure}
\begin{center}
\epsfig{file=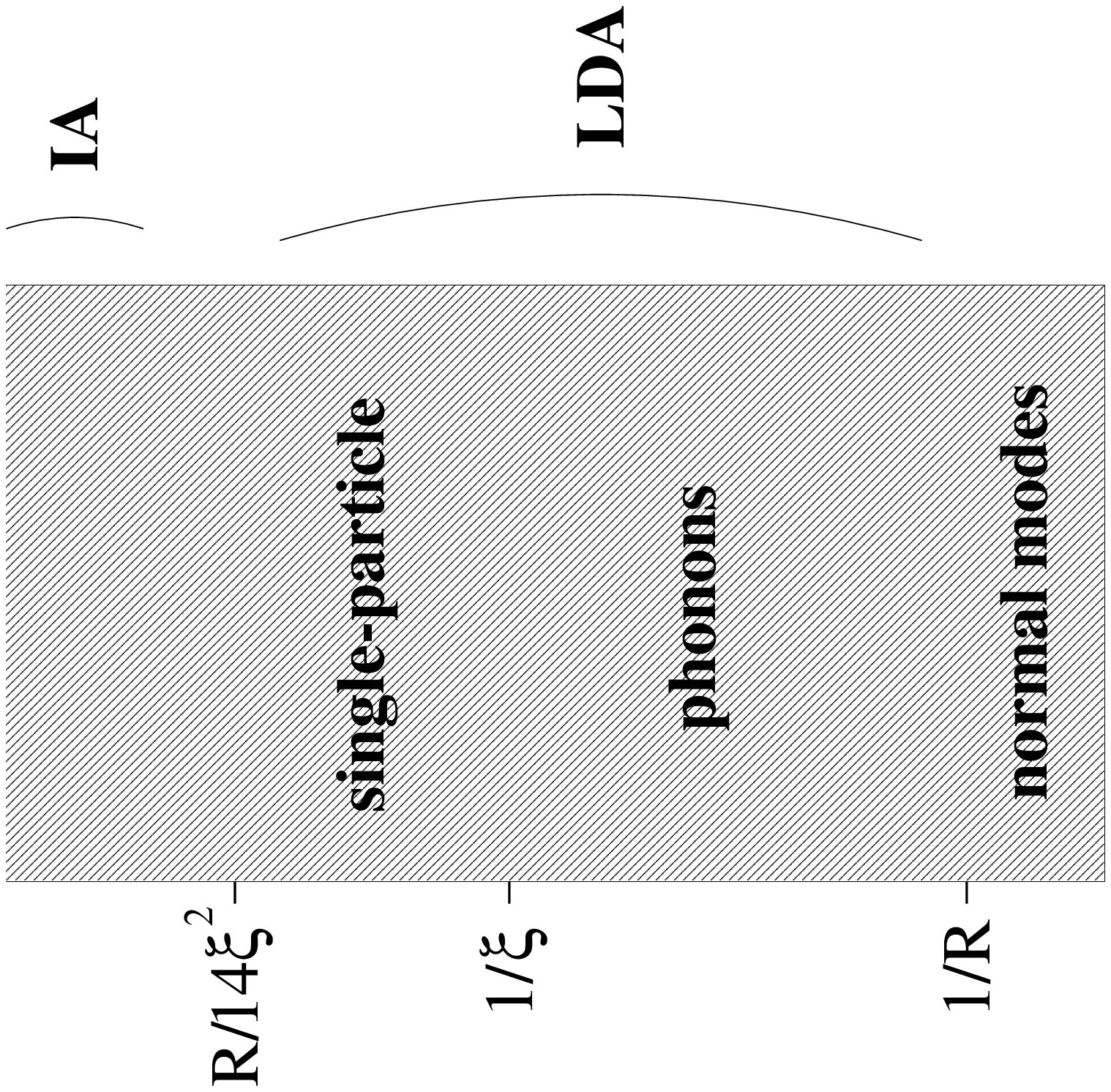, width=0.75\linewidth, angle=-90}
\begin{caption}
{Relevant wave vector regimes characterizing the dynamic
structure factor at $T=0$. $R$ is the size of the condensate
(\ref{RTF}); $\xi$ is the healing length (\ref{healing}). 
The ranges of applicability of the local density approximation (LDA)
and of the impulse approximation (IA) are also schematically
indicated. The momentum transfer $\hslash R/14\xi^2$ corresponds to
the condition $\Delta_{\rm LDA}=\Delta_{\rm IA}$ for the LDA and IA
widths of the dynamic structure factor (see Eq.(\ref{wratio})).}
\end{caption}
\label{fig11}
\end{center}
\end{figure}


\begin{references}
\bibitem{Palewski}
H.~Palevski, K.~Otnes, and K.E.~Larson, Phys. Rev. \textbf{112}, 11 (1958);
D.G.~Henshaw and A.D.B.~Woods, Phys. Rev. \textbf{121}, 1266 (1961).

\bibitem{Sokol} P.~Sokol, In: \textit{Bose-Einstein Condensation},
Ed. A.~Griffin, D.W.~Snooke, and S.~Stringari, 
p. 51. Cambridge University Press, 1995.

\bibitem{VanHove}
L.~Van~Hove, Phys. Rev. \textbf{95}, 249 (1954).

\bibitem{Bragg1} J.~Stenger, S.~Inouye, A.P.~Chikkatur, D.M.~Stamper-Kurn,
D.E.~Pritchard, W.~Ketterle, Phys. Rev. Lett.
\textbf{82}, 4569 (1999).

\bibitem{Bragg2}
D.M.~Stamper-Kurn, A.P.~Chikkatur, A.~G\"orlitz, S.~Inouye,
S.~Gupta, D.E.~Pritchard, and W.~Ketterle, 
Phys. Rev. Lett. {\bf 83}, 2876 (1999).

\bibitem{IG}
A.L.~Fetter and J.D.~Walecka, \textit{Quantum theory of many-particle
systems}, (McGraw-Hill, New York, 1971).

\bibitem{Hohenberg} P.C.~Hohenberg and P.M.~Platzman, Phys. Rev.
\textbf{152}, 198 (1966).

\bibitem{LP}
L.P.~Pitaevskii, Sov. Phys. JETP \textbf{13}, 451 (1961).

\bibitem{Fetter1}
A.L.~Fetter, Ann. Phys. (N.Y.) \textbf{70}, 67 (1972).

\bibitem{Griffin}
Wen-Chin~Wu and A.~Griffin, Phys. Rev. A \textbf{54}, 4204 (1996).

\bibitem{Bogoliubov}
N.~Bogoliubov, J. Phys. (Moscow) \textbf{11}, 23 (1947).

\bibitem{Timmermans}
E.~Timmermans and P.Tommasini, e-print cond-mat/9707322.

\bibitem{rmp} F.~Dalfovo, S. Giorgini, L. Pitaevskii and S. Stringari,
Rev. Mod. Phys. \textbf{71}, 463 (1999).

\bibitem{notaIG}
In the presence of harmonic trapping the static structure factor of
an ideal Bose gas is actually given by $S(q)=
1-\exp{[q^2/(\hslash m\omega_z)]}$, and vanishes when $q\to 0$. For the
values of $q$ employed in the experiment of \cite{Bragg2} the prediction
of the ideal Bose gas model is however always close to 1.

\bibitem{Huang}
K.~Huang, \textit{Statistical Mechanics}, 2nd ed. (Wiley, New-York, 1978).

\bibitem{Baym}
G.~Baym and C.~Pethick, Phys. Rev. Lett. \textbf{76}, 6 (1996).

\bibitem{Franco}
F.~Dalfovo, L.~Pitaevskii, and S.~Stringari, Physica Scripta
T\textbf{66}, 234-237 (1996).

\bibitem{Franco2}
F.~Dalfovo, L.~Pitaevskii, and S.~Stringari, Phys. Rev. A
\textbf{54}, 4213 (1996).

\bibitem{Lundh}
E.~Lundh, C.J.~Pethick, and H.~Smith, Phys. Rev. A
\textbf{55},  2126 (1997).

\bibitem{notacohe}
The coherence length should not be confused with the healing
length $\xi$ (\ref{healing}) which, differently from $\chi_x$,
becomes small as the density of the sample increases.

\bibitem{Landauqm}
L.D.~Landau and E.M.~Lifshitz, \textit{Quantum Mechanics}, $\S$ 131,
3rd edition, (Pergamon Press, Oxford 1977).

\bibitem{notaeiko}
The eikonal approximation has been successfully used for
describing both electron-nucleus scattering and neutron scattering from liquid
helium (see \cite{Rinat,Carraro} and references therein).

\bibitem{Danthesis}
D.M.~Stamper-Kurn, PhD. Thesis (1999).

\bibitem{Lipparini}
E.~Lipparini and S.~Stringari, Phys. Rep. \textbf{175}, 3 \& 4,
103-261 (1989).

\bibitem{Feenberg}
E.~Feenberg, {\it Theory of Quantum fluids} (Academic, New York, 1969),
Chap. 4; D.~Hall and E.~Feenberg, Ann. Phys. \textbf{63},
335 (1971).

\bibitem{Pines}
D. Pines and P. Nozi\`eres, {\it The Theory of Quantum Liquids}
(Benjamin, New York, 1966) Vol. I.

\bibitem{Sick}
D.~Day, J.S.~McCarthy, T.W.~Donnelly, and I.~Sick,
Annu. Rev. Nucl. Part. Sci. \textbf{40}, 357 (1990).

\bibitem{Sears}
V.F.~Sears, Phys. Rev. B \textbf{30}, 44 (1984).

\bibitem{Carraro}
C.~Carraro and S.E.~Koonin, Nucl. Phys. \textbf{A524}, 201 (1991).

\bibitem{vortexJILA}
M.R.~Matthews, B.P.~Anderson, P.C.~Haljan, D.S.~Hall, C.E.~Wieman,
and E.A.~Cornell, Phys. Rev. Lett. {\bf 83}, 2498 (1999).

\bibitem{vortexENS}
K.W.~Madison, F.~Chevy, W.~Wohlleben, and J.~Dalibard, e-print
cond-mat/9912015.

\bibitem{Michele}
E.~Lundh, C.J.~Pethick, and H.~Smith, Phys. Rev. A
\textbf{58},  4816 (1998);
F.~Dalfovo and M.~Modugno, e-print cond-mat/9907102, to appear
in Phys. Rev. A.

\bibitem{Francesca}
F.~Zambelli and S.~Stringari, Phys. Rev. Lett.
\textbf{81}, 1754 (1998);
A.A.~Svidzinsky and A.~Fetter,  Phys. Rev. A
\textbf{58}, 3168 (1998).

\bibitem{dislocazione}
E.L.~Bolda and D.F.~Walls, Phys. Rev. Lett.
\textbf{81}, 5477 (1998).

\bibitem{GP} E.P.~Gross, Nuovo Cimento \textbf{20}, 454 (1961).

\bibitem{GPnum}
F.~Dalfovo and S.~Stringari, Phys. Rev. A \textbf{53}, 2477 
(1996).

\bibitem{MIT} M.R.~Andrews, C.G.~Townsend, H.-J.~Miesner,
D.S.~Durfee, D.M.~Kurn, W.~Kettele,
Science \textbf{275}, 637 (1997).

\bibitem{Lev}
L.P.~Pitaevskii and S.~Stringari, Phys. Rev. Lett. \textbf{83}, 4237
(1999).

\bibitem{Rinat}
S.A.~Gurvitz and A.S.~Rinat, Phys. Rev. C \textbf{47}, 2901
(1993).

\end{references}
\end{document}